\documentclass[
    aps,
    prx,
    amsmath,amssymb,
    tightenlines, 
    superscriptaddress,
    floatfix,
    reprint,
]{revtex4-2}

\usepackage[british]{babel} 
\usepackage[utf8x]{inputenc} 

\usepackage{newtxtext,newtxmath} 
\usepackage{microtype} 

\usepackage{bm}
\usepackage{mathtools} 
\usepackage{physics} 
\usepackage{dsfont} 
\usepackage{braket} 
\usepackage{stackengine} 
\usepackage{lipsum} 
\usepackage{amsmath} 

\usepackage[cal=cm,scr=dutchcal]{mathalpha} 

\usepackage{graphicx}
\usepackage[svgnames]{xcolor}
\usepackage{siunitx} 

\usepackage{ragged2e} 
\usepackage{array} 
\usepackage{tabularx} 
\usepackage{booktabs} 
\usepackage[normalem]{ulem} 
\usepackage{placeins}

\definecolor{cset-aps-blueberry}{RGB}{28,128,158}
\definecolor{cset-aps-blue}{RGB}{46,44,184}
\definecolor{cset-aps-turquoise}{RGB}{0,67,88}
\definecolor{cset-aps-limegreen}{RGB}{190,219,67}
\definecolor{cset-aps-green}{RGB}{31,138,112}
\definecolor{cset-aps-yellow}{RGB}{255,225,25}
\definecolor{cset-aps-orange}{RGB}{253,116,0}
\definecolor{cset-aps-red}{RGB}{219,0,43}

\definecolor{cset-aps-kobalt-medium}{RGB}{62,54,222}
\definecolor{cset-aps-kobalt-dark}{RGB}{28,24,150}

\definecolor{cset-aps-my-label-red}{RGB}{202,0,17}
\definecolor{cset-aps-my-label-blue}{RGB}{53,71,140}
\definecolor{cset-aps-my-label-gray}{RGB}{145,145,145}

\usepackage{hyperref}
\hypersetup{%
    colorlinks=true,
    linkcolor={cset-aps-blue},
    linkbordercolor={cset-aps-red},
    filecolor={cset-aps-orange},
    filebordercolor={cset-aps-orange},
    citecolor={cset-aps-kobalt-medium},    
    citebordercolor={cset-aps-kobalt-medium},
    urlcolor={cset-aps-kobalt-dark},
    urlbordercolor={cset-aps-kobalt-dark},
    menucolor={cset-aps-limegreen},
    menubordercolor={cset-aps-limegreen},
    breaklinks=true,
    pdfborderstyle={/S/U/W 2},
    pdfpagemode=UseOutlines,
    pdfstartpage={1}
}
\usepackage{enumitem} 

\newcommand{\pss}{\ensuremath{/\mathrm{s}^2}} 
\newcommand{\amp}{A} 
\newcommand{\base}{B} 
\newcommand{\mf}{{m_F}} 
\newcommand{\rb}{$^{87}$Rb} 
\renewcommand{\qty}[2]{#1\,\text{#2}} 
\newcommand{\ie}{i.\,e.}
\newcommand{\eg}{e.\,g.}
\newcommand{\cf}{cf.}
\newcommand{\all}{\text{all}}
\newcommand{\summ}{\text{sum}}
\newcommand{\diff}{\text{diff}}
\newcommand{\laser}{\text{las}}
\newcommand{\offset}{\text{off}}
\newcommand{\recons}{\text{rec}}
\newcommand{\bias}{\text{bias}}
\renewcommand{\set}{\text{set}}
\newcommand{\sett}{\text{set}}
\newcommand{\eff}{\text{eff}}

\newcommand{\ext}{\text{ext}}
\newcommand{\mm}{Materials and Methods}
\newcommand{\supp}{Supplementary Materials}

\newcommand{\sgn}{\operatorname{sgn}}

\newcommand{\orcid}[1]{\href{https://orcid.org/#1}{\includegraphics[width=7pt]{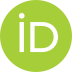}}}

\newcommand{\affTUDa}{\href{https://ror.org/05n911h24}{Technische Universit{\"a}t Darmstadt}, Fachbereich Physik, Institut f{\"u}r Angewandte Physik, Schlo{\ss}gartenstra{\ss}e 7, 64289 Darmstadt, Germany}
\newcommand{\affHFHF}{Helmholtz Forschungsakademie Hessen f{\"u}r FAIR (HFHF), \href{https://ror.org/02k8cbn47}{GSI Helmholtzzentrum f{\"u}r Schwerionenforschung}, 64291 Darmstadt, Germany}

\allowdisplaybreaks 

\AtBeginDocument{%
}

\begin{document}

\title[Parameter Estimation via Amplitude Collapse]{
Parameter Estimation from Amplitude Collapse in Correlated Matter-Wave Interference
}

\author{Daniel Derr\,\orcid{0000-0002-8690-3897}}
\email{daniel.derr@physik.tu-darmstadt.de}
\email{daniel.derr@gmx.net}
\thanks{These authors contributed equally to this work.}

\author{Dominik Pfeiffer\,\orcid{0000-0003-4530-3251}}
\email{apqpub@physik.tu-darmstadt.de}
\thanks{These authors contributed equally to this work.}

\affiliation{\affTUDa}
\author{Ludwig Lind\,\orcid{0009-0000-3354-511X}}
\affiliation{\affTUDa}

\author{Gerhard Birkl\,\orcid{0000-0002-4137-9227}}
\affiliation{\affTUDa}
\affiliation{\affHFHF}

\author{Enno Giese\,\orcid{0000-0002-1126-6352}}
\affiliation{\affTUDa}

\begin{abstract}
Operating matter-wave interferometers as quantum detectors for fundamental physics or inertial sensors with unprecedented accuracies relies on noise rejection, often implemented by correlating multiple sensors.
They can be spatially separated (gradiometry or gravitational-wave detection) or consist of different internal states (magnetometry or quantum clock interferometry), with a signal-amplitude modulation serving as a signature of a differential phase.
In this work, we introduce Parameter Estimation from Amplitude Collapse (PEAC) by applying statistical inference techniques for different magnetically sensitive substates of an atom interferometer.
We demonstrate that PEAC provides higher trueness, resulting in a substantially reduced bias compared to standard methods for perfectly correlated signals, while achieving competitive precision near, but not at, vanishing amplitudes.
This indicates that vanishing signals do not constitute the most favourable working point for high-accuracy sensing, relevant to quantum clock interferometry.
PEAC presents a generally applicable complementary evaluation method for correlated interferometers without phase stability, increasing the overall accuracy and enabling applications beyond atom-based interferometry.

\vspace{2mm}\noindent This article has been published in \href{https://doi.org/10.1126/sciadv.aeg5334}{Science Advances \textbf{12}, aeg5334 (2026)} under the terms of the \href{https://creativecommons.org/licenses/by/4.0/}{Creative Commons Attribution License 4.0 [CC BY]}.
DOI: \href{https://doi.org/10.1126/sciadv.aeg5334}{10.1126/sciadv.aeg5334}
\end{abstract}

\maketitle

\section{INTRODUCTION}
To exploit the full potential of atom interferometers as quantum sensors at the highest accuracy, the signal of interest is often encoded in the differential phase $\theta$ between two correlated interferometers.
This strategy suppresses common-mode noise and is an established, yet crucial tool for high-precision measurements, \eg{} setting new records in the determination of the fine-structure constant~\cite{morel:2020:determination,richard:2018:measurement}, forming the working principle of quantum-based gravitational-wave or dark-matter detectors~\cite{arvanitaki:2018:search, derr:2023:clock, abend:2024:terrestrial} on large baselines~\cite{abdalla:2025:terrestrial, abend:2024:terrestrial}, and being central to operational and proposed quantum tests of fundamental physics~\cite{rosi:2014:precision,lamporesi:2008:determination,fixler:2007:atom, pikovski:2015:uni_decoh} and general relativity~\cite{schlippert:2014:quantum,asenbaum:2020:atominterferometric,barret:2016:dual,schaffrath:2025:unified,yu:2011:gravitational,chaibi:2016:lowfrequency,barret:2015:correlative,barret:2022:testing}.
It is also key for robust sensors~\cite{bongs:2019:taking,elliot:2023:quantum} used in magnetometry~\cite{meister:2025:space,hardman:2016:simultaneous,zhou:2010:precisely,zhong-kun:2011:simultaneous}, gyroscopy~\cite{gustanvson:2000:rotation,berg:2015:composite,canuel:2006:sixaxis,beydler:2024:guidedwave}, or gradiometry~\cite{sorrentino:2014:sensitivity,snadden:1998:measurement,bertoldi:2006:atom} with applications in inertial sensing, exploration, and navigation~\cite{canuel:2006:sixaxis,dickerson:2013:multiaxis}.
In such real-world applications, the differential phase can be extracted even in noisy environments, where the interference signal cannot be resolved in a single interferometer.

However, differential phase estimation in correlation measurements poses its own challenges even for established techniques like fitting ellipses to parametric plots of noisy, but correlated bivariate data which extract the phase from the ellipses' eccentricity~\cite{foster:2002:method,zhang:2023:dependence,fitzgibbon:1999:direct,szpak:2012:guaranteed}.

Since concepts for specifying measurement quality are not used consistently across subfields, we adopt the following definitions~\cite{vim:2012:defis, iso:2023:accuracy}:
\emph{Trueness} denotes the closeness of the estimated (reconstructed) phase $\theta_\recons$ to the true (\eg{} set) phase $\theta_\set$.
As a practical quantitative measure of trueness, we consider the bias $\theta_\bias =\theta_\recons-\theta_\set$.
\emph{Precision} characterises the closeness of agreement of repeated estimates under identical conditions.
We therefore choose the spread, \ie{} the standard deviation associated with the estimation process, as a quantifiable measure.
\emph{Accuracy} refers qualitatively to the combined effect of trueness and precision.
While typically \emph{precise}, ellipse-based phase estimation becomes biased except for a circle ($\theta = \pi/2$ and odd multiples).
These are the only phases where the estimator is unbiased.
For all other cases, a systematic error is introduced.
The \emph{trueness} of the phase estimate degrades strongly near degeneracy points ($\theta = \pi \mathds{Z}$), where the underlying ellipse collapses to a line.

In this work, we propose and implement Parameter Estimation from Amplitude Collapse (PEAC) as an additional, complementary statistical technique for noisy matter-wave interferometers, and implement it for correlated interferometers in magnetometry as well as for velocimetry by deliberately inducing an amplitude collapse.
We demonstrate, both experimentally and theoretically that PEAC enhances the trueness by a bias reduction for a broad range of phase settings in magnetometry, albeit at the cost of precision, yielding overall improved accuracy also near the degeneracy points, \ie{} for perfectly in-phase or $\pi$-phase-shifted correlated interferometers.

PEAC is inspired by quantum clock interferometry~\cite{roura:2020:gravitational,dipumpo:2021:gravitational,rosi:2017:quantum, borregaard:2025:qclock, fromonteil:2025:nonlocal, guendogan:2026:qclock}, where a superposition of internal states undergoes an atom-interferometer sequence, creating a spatial superposition in which possible relativistic~\cite{zych:2011:quantum, covey:2025:qclock_curvature} and proper-time effects~\cite{loriani:2019:interference} encode a periodic signal-amplitude loss associated with which-path information~\cite{margalit:2015:selfinterfering}.
Here, the internal states acquire different interferometer phases, for example due to relativistic effects associated with mass-energy equivalence, but are not measured individually.
In particular, atoms in different internal states possess slightly different rest masses due to their internal energy, known as mass defect~\cite{yudin:2018:mass,asano:2024:quantum}, which leads to state-dependent phase evolution in the interferometer.
Although these differential phase shifts are typically minute and can, in certain situations, also be extracted using the methods discussed below, the question remains how such a signature can be most effectively utilised in quantum clock interferometry.
Recent experiments~\cite{zhou:2025:geometric} operate beyond the perturbative regime, with differential phases induced by magnetic field gradients, a mechanism that we also employ in the present work.

Similar to those quantum-clock schemes, PEAC does not necessarily require measuring correlated interferometers individually, but relies on the detection of (possibly incoherent) mixtures of them.
A differential phase shift between the individual interferometers induces a beating of their superposed interference fringes.
This characteristic structure manifests as collapses and revivals of the combined signal's oscillation amplitude, whose pattern appears in statistical histograms of population measurements, extractable~\cite{dickerson:2013:multiaxis,geiger:2011:detecting,abend:2016:atomchip,pelluet:2025:atom} even when suffering large phase noise.

We analyse the PEAC performance in atom interferometers containing a stochastic mixture of magnetically sensitive substates in order to estimate the acceleration induced by a magnetic-field gradient.
Using state-selective measurements, we compare this technique to corresponding ellipse fits, demonstrating that PEAC provides a complementary statistical analysis of existing correlation data not through joint phase fitting, but via an optimally chosen transformation of the bivariate coordinates and their amplitude marginal distributions.
Such an analysis can be always performed and optimised in addition to ellipse fits.
PEAC demonstrates improved overall performance, with the highest accuracy for phases near, but not exactly at degeneracy points, in contrast to expectations from Gaussian uncertainty propagation.
In fact, the signal amplitude is small but still finite, providing a working point that exhibits negligible bias and precision comparable to ellipse fitting.
This observation highlights a distinct operating regime that differs from geometric phase amplification~\cite{zhou:2025:geometric}.

The method extracts a measurand, in our case the differential phase of interest, from histograms of a suitably chosen linear combination of correlated signals.
It therefore requires a model for the dominant noise and for the functional dependence of the probability distribution on the measurand, in our case through the amplitude of an interference pattern.
The method also depends on the weights or mixture fractions that define the combined signal.
These weights, or other distribution-specific parameters, may be determined either by scanning the measurand or, if this is experimentally inaccessible, from independent experiments, as demonstrated below.

PEAC applies to a wide range of cases: coherent implementations like quantum clocks as well as incoherent, stochastic mixtures of components.
It allows extracting phase information in setups previously considered unsuitable, for example, when these components spatially overlap.
At the same time, it can also be applied in situations where components can be individually resolved, a necessary condition for ellipse fitting.
Moreover, PEAC enables phase extraction even when phase fluctuations prevent fringe resolution in correlated interferometer signals~\cite{zheng:2023:labbased,bothwell:2022:resolving} and when the differential phase cannot be tuned, providing a valuable toolbox for next-generation platforms~\cite{meister:2025:space,barret:2016:dual,elliot:2023:quantum,pelluet:2025:atom,geiger:2011:detecting,struckmann:2024:platform} for matter-wave sensors with ultracold quantum gases.
Furthermore, we extend the approach to related estimation problems in which information is encoded in contrast or envelope parameters, such as the extraction of effective temperatures or rotation rates from contrast-envelope fitting in atom interferometry~\cite{mueller:2008:atom, kovachy:2015:quantum, castanet:2024:atom}.
Last, it can be combined with any statistical inference technique, such as least-squares fitting or maximum likelihood estimation.

\section{RESULTS}

\subsection{PEAC Methodology}
\label{sec:PEAC}
PEAC is an estimation method applicable to interferometric experiments that extracts the value of a parameter of interest from the amplitude of an interference signal without the need to resolve an individual interference fringe.
We demonstrate this for two distinct parameters:
the differential phase between correlated interferometers, extracted from the characteristic amplitude collapse of a combined signal for magnetometry in Sect.~\ref{subsubsec:magnetometry}, and the momentum width of an atomic cloud, extracted from the decreasing amplitude upon deliberately opening the interferometer to reduce the clouds' overlap in Sect.~\ref{subsubsec:velocimetry}.

The parameter of interest is extracted from a statistical model with a known probability density function (PDF), which is experimentally probed by repeated measurements used to construct histograms.
The PDF can be fitted directly to these histograms, or one can use more sophisticated procedures such as maximum likelihood estimation to extract the PDF's parameters (\cf{} \supp).

The PDF generally depends not only on the parameter of interest but also on additional experimental quantities, such as the intrinsic contrast.
If a control parameter is experimentally accessible that induces a characteristic variation of the amplitude, PEAC can be operated in scanning mode, permitting the simultaneous extraction of both intrinsic parameters and the quantity of interest.
We demonstrate this mode of operation by scanning the differential phase for magnetometry and the timing asymmetry for velocimetry, from which the respective quantities of interest are extracted via a fit to the amplitude model.

If no such control parameter is accessible, but the intrinsic parameters of the PDF are sufficiently characterised by independent measurements, PEAC is operated in static mode, allowing the quantity of interest to be estimated from a single experimental configuration.
We demonstrate static operation in Sect.~\ref{sec:correlation}, where we independently determine the intrinsic contrast from the individual signals of two correlated interferometers, which is then used to infer the magnetic field gradient without the necessity to scan the differential phase.

\subsubsection{Magnetometry}
\label{subsubsec:magnetometry}
\begin{figure}[!h]
    \centering
    \includegraphics[scale=.9]{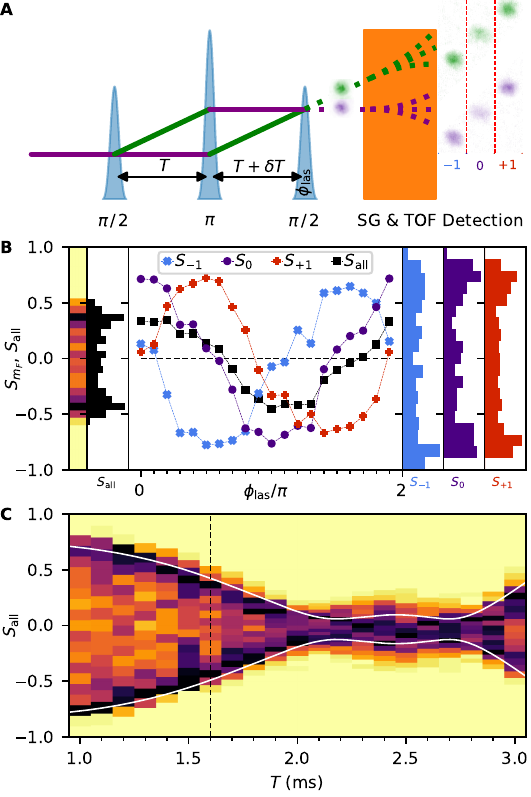}
    \caption{\textbf{Spacetime diagram and signal extraction.}
        (\textbf{A}) Mach--Zehnder atom interferometer (MZI) realised by three Bragg pulses separated by $T$ and $T+\delta T$, respectively, driving transitions between momentum states $\ket{0 \hbar k_\eff}$ (purple) and $\ket{1 \hbar k_\eff}$ (green).
        The phase of the final pulse is shifted by $\phi_\laser$, thus scanning the interference fringes at the two exit ports.
        The MZI is followed by a time of flight (TOF) that spatially separates the exits, during which an optional Stern--Gerlach (SG) field can separate the $\mf$ substates of the $F=1$ ground state within each momentum class in addition.
        (\textbf{B}) Fringe scans ($20$ settings of $\phi_\laser$, $15$ repetitions averaged, uncertainties omitted for visual clarity) at $T = \qty{1.6}{ms}$.
        Without the SG field, we observe the fringe $S_\all$ (black), with the corresponding histogram from $300$ measurements (left marginal, also displayed as a density plot on the far left) has a double-peak structure that depends on the signal's amplitude.
        Applying an SG field allows resolving the individual signals $S_\mf$ per substate (blue, indigo, red), whose histogram distributions are shown at the right marginal.
        The individual signals show equal amplitudes, while the amplitude of $S_\all$ is reduced due to summation of the individual $S_\mf$ with shifted phases.
        (\textbf{C}) The combined density plot of the histograms of $S_\all$, obtained for a scan of the interferometer time $T$, reveals the collapse and revival of the amplitude caused by the incoherent overlay of the three signals $S_\mf$.
        The white lines trace the calculated amplitude as $\pm A_\all$, parametrised by Eq.~\eqref{eq:A_all}.
        The black dashed line corresponds to the data shown in (\textbf{B}).
        }
    \label{fig:1}
\end{figure}

In this subsection, we illustrate PEAC in scanning mode for a concrete experimental implementation exploiting the connection between the beat amplitude and a differential phase in a three-level system, as well as the associated level populations.
Specifically, we apply PEAC to atom interferometry-based magnetometry~\cite{meister:2025:space,hardman:2016:simultaneous,zhou:2010:precisely,zhong-kun:2011:simultaneous} in one dimension, a benchmark application well-suited to typical vibration-prone environments and compact devices, even when state-selective interferometry is unavailable.
Ultracold \rb{} atoms from a Bose--Einstein condensate (BEC) provide a suitable point source~\cite{dickerson:2013:multiaxis} with sub-recoil momentum width, coherence, and access to efficient Bragg diffraction~\cite{hartmann:2020:regimes}.
Our experiment~\cite{lauber:2011:pra} creates BECs in a far-detuned crossed optical dipole trap (CDT), typically with $20\,000$ atoms at $\qty{20(5)}{nK}$ and a condensate fraction exceeding $\qty{80}{\%}$.
When no optical pumping or magnetic quantisation is applied, the atoms occupy a stochastic mixture of all magnetic substates $\mf \in \{-1,0,+1\}$ of the ground-state manifold $ \ket{5^2S_{1/2}, F=1}$.
All states are addressed and Bragg-diffracted by a pair of counter-propagating laser beams~\cite{pfeiffer:2025:dichroic} with a frequency difference $\Delta \omega  = 2\pi \times \qty{15.084}{kHz}$, coupling the two momentum states $\ket{0 \hbar k_\eff}$ and $\ket{1 \hbar k_\eff}$ with the modulus $ k_\eff = 2 \times 2\pi / \qty{780.226}{nm}$ of the effective wave vector of the two-photon Bragg transition.
With these Bragg pulses, we implement a Mach--Zehnder atom interferometer (MZI) in a $\pi/2$--$\pi$--$\pi/2$ pulse configuration as illustrated in Fig.~\ref{fig:1}(\textbf{A}).
The laser-intensity envelopes of all three pulses are shaped as Blackman window pulses~\cite{blackman:1958:measurement,weisstein:blackman} of length $\tau = \qty{100}{\textmu s}$.
We label the temporal separation of the maxima of two consecutive pulses by $T$ and $T+\delta T$, respectively.
Identical values $T \in [\qty{1}{ms}, \qty{3}{ms}]$ and $\delta T =0$ result in a total duration of the MZI sequence up to $2T+\tau = \qty{6.1}{ms}$.

After the symmetric MZI, a time of flight (TOF) spatially separates the interferometer output ports associated with the momentum states $\ket{j \hbar k_\eff}$ for measuring the atom numbers $N_j(\phi) = \base_{j} \pm \amp_{j} \cos\phi$ in different momentum classes $j\in\{0,1\}$, with the interferometric phase $\phi$, baseline $\base_j$ and amplitude $\amp_j$.
We define the interferometer signal as the normalised population difference, \ie{}
\begin{equation}
\label{eq:def_S}
    S_\all (\phi)  =\frac{N_0(\phi) - N_1(\phi)}{N_0(\phi) + N_1(\phi)} \cong \base_\all +\amp_\all \cos \phi,
\end{equation}
where the signal's baseline $\base_\all$ and amplitude $\amp_\all$ satisfy $\lvert \base_\all \rvert + \lvert \amp_\all \rvert \le 1$ and depend on $\base_j, \amp_j$.
For more details on the experimental apparatus and the derivation of $S_\all$ we refer to \mm.

By characterising $S_\all$, for example by scanning an interferometer fringe, one can extract the interferometric phase $\phi$ and thus the quantity of interest influencing $\phi$.
For instance, this could be the projection $a_\ext$ of an external acceleration in the direction of the Bragg beams.
To scan an interference fringe, we offset the phase of the final $\pi/2$ pulse by $20$ evenly spaced values of $\phi_\laser \in [0, 2\pi)$.
We perform $15$ repetitions per $T$ for improved signal to noise.
A typical fringe for $T = \qty{1.6}{ms}$ is shown in the centre panel of Fig.~\ref{fig:1}(\textbf{B}) by the black trace.
We see a clear oscillation and periodicity of the fringe characterised by its amplitude $\amp_\all$, phase $\phi$, and baseline $\base_\all$.
If $\phi$ remains stable between all pulses, a cosine can be fitted to the fringe, providing direct access to these parameters.
Conversely, in the absence of phase stability, the amplitude $\amp_\all$ and baseline $\base_\all$ of $S_\all$ can be obtained via a statistical analysis~\cite{geiger:2011:detecting, chiow:2011:102, berrada:2013:integrated, pelluet:2025:atom}, albeit at the cost of losing the phase information.
This procedure also can be applied to data obtained under phase-stable conditions, such as our measurements at small $T$.

To demonstrate this statistical method, we show, at the left margin of Fig.~\ref{fig:1}(\textbf{B}), a histogram associated with the fringe in black, where the statistics is generated from all $300$ contributing measurements.
The very left column shows the same histogram converted into a density plot using false-colour coding.
We observe the typical double-peak structure of the histogram~\cite{geiger:2011:detecting,chiow:2011:102}, where the distance of the peaks is roughly given by $2 \amp_\all$ and their individual width is determined by baseline fluctuations of $\base_\all$.
This characterisation is valid for a homoscedastic signal, where $\amp_\all$ does not fluctuate, as specified in \mm{}, which is dominated by technical noise and operates well above the quantum-projection noise limit.

In addition to scanning $\phi$ via $\phi_\laser$ for a fixed value of $T$, we also vary the interferometer time $T \in [\qty{1}{ms}, \qty{3}{ms}]$ in steps of $\qty{100}{\textmu s}$.
The corresponding histograms are displayed as a composite density plot in Fig.~\ref{fig:1}(\textbf{C}), where each column is obtained in the same way as in Fig.~\ref{fig:1}(\textbf{B}) corresponding to the histogram behind the dashed black line at $T = \qty{1.6}{ms}$.
We observe a decrease of the signal amplitude for increasing $T$ with a collapse at $T = \qty{2.1}{ms}$ which is not due to loss of coherence, since $\amp_\all$ revives for longer times.
This behaviour is caused by an underlying modulation of $\amp_\all(T)$, which is a consequence of the beat of the three signals $S_\mf (\phi_\mf)$ defined in analogy to Eq.~\eqref{eq:def_S} and associated with three $\mf$ substates.
The $\mf$ substates experience a state-dependent acceleration, in part due to a magnetic field gradient, acquiring different phases $\phi_\mf$, which induce this characteristic collapse and revival behaviour.
The magnetically insensitive $\mf=0$ substate acquires the phase $\phi_0$, including $\phi_\laser$ and accelerations and perturbations common to all substates.
The magnetically sensitive states with $\mf \in \{+1,-1\}$ pick up the phases $\phi_{\pm1} = \phi_0 \pm \theta/2$ with the differential phase
\begin{equation}
\label{eq:theta}
    \theta \cong 2 \, k_\eff a_\ext T^2 (1 + 0.1486 \, \tau/T )
\end{equation}
between the two substates, induced by a Zeeman force that accelerates the $\mf=+1$ state by $a_\ext$ in the direction of the Bragg beams and the $\mf=-1$ state by the same magnitude but in the opposite direction.
Here, we correct the conventional MZI phase for finite-pulse durations~\cite{bertoldi:2019:phase} in $\tau/T$, which can approach $0.1$ in our case with $\tau=\qty{100}{\textmu s}$ for short interferometer times.
We present the derivation of the numerical factor in \mm.

In our experiment, we have additional access to the individual substates by applying the Stern--Gerlach (SG) method~\cite{gerlach:1989:sterngerlach, bauer:2023:sterngerlach} during a TOF after the MZI:
In each momentum exit port we can create a spatial separation of the substates by applying an additional magnetic field gradient as depicted in Fig.~\ref{fig:1}(\textbf{A}) at right.
This method allows us to determine the oscillatory signal $S_{\mf}$ of the individual substates as displayed in the centre panel of Fig.~\ref{fig:1}(\textbf{B}), along with the corresponding histograms at the right marginals.
As expected, we observe the same oscillation amplitude $\amp_{\mf}$ for all three substates, which is in particular larger than $\amp_\all$.
Based on this insight, we conclude that in a non-state-selective setting, the observable is $S_\text{all} = \lambda_{+1} S_{+1} + \lambda_0 S_0 + \lambda_{-1} S_{-1}$, where $\lambda_{\mf}$ are the weights of each substate in the incoherent stochastic mixture.

In the remainder of this section, we focus on this non-state-selective setting.
Here, the observable $S_\all$ can be written as $S_\all = \base_\all +\amp_\all \cos (\phi_0 + \theta_\offset)$.
The offset $\theta_\offset$ can be interpreted as a geometric phase~\cite{zhou:2025:geometric} and depends on $\theta$ and $\lambda_\mf$ (\cf{} \mm).
Here, $\base_\all = \sum\displaylimits_{\mf} \lambda_\mf \base_\mf$ is the baseline, and
\vspace{-1.5mm}
\begin{equation}
\label{eq:A_all}
    \amp_\all(\theta) = \amp_0 \sqrt{\Delta \lambda^2 \sin^2 (\theta/2) + \bigl[\lambda_0 + (\lambda_{+1} + \lambda_{-1}) \cos (\theta/2) \bigr]^2}
\end{equation}
is the amplitude, with the population imbalance $\Delta \lambda = \lambda_{+1} - \lambda_{-1}$, assuming $\amp_{-1} = \amp_0 = \amp_{+1}$.
In this particular setting, we cannot access the differential phase $\theta$ directly from a histogram recorded for a single interferometer time $T$, since the weights $\lambda_\mf$ are unknown.
However, scanning the interferometric time $T$ allows to infer $\theta$.
The underlying reason is the characteristic variation of the amplitude $\amp_\all$ with $T$, as evident from Eqs.~\eqref{eq:theta} and \eqref{eq:A_all}, due to the overlay of the three individual interferometer signals with their respective phases.
Crucially, the technique does not intrinsically require a scan of $\theta$, provided the auxiliary parameters $A_0$ and $\lambda_\mf$ are constrained by independent measurements or deterministic tuning, as is done in quantum clock experiments.
In this case, the static mode of PEAC can be used to reconstruct the phase for each experimental setting.

In the present section, PEAC in its scanning mode is explicitly performed by fitting the histogram's probability density function (PDF) for each $T$ to obtain $\amp_\all(T)$, where the form of the PDF~\cite{geiger:2011:detecting, berrada:2013:integrated, pelluet:2025:atom} and further details can be found in \mm.
Using $\amp_\all(T)$, we fit Eq.~\eqref{eq:A_all} with the help of Eq.~\eqref{eq:theta} to extract the acceleration and further parameters $a_\ext = \qty{32.2(1)}{mm\pss}$, $\lambda_0=\qty{0.42(1)}{}$, $\Delta\lambda=\qty{0.18(1)}{}$, and $\amp_0 = \qty{0.79(1)}{}$.
We calculate the modulated amplitude $\amp_\all(T)$ from these values and plot $\pm \amp_\all(T)$ as white lines in Fig.~\ref{fig:1}(\textbf{C}).
The uncertainty of $a_\ext$ and all other experimental uncertainties in the following are obtained via bootstrapping~\cite{efron:1994:introduction,efron:2000:bootstrap,zoubir:1998:bootstrap}, as explained in \mm.

Even in the presence of phase instability and a non-state-selective setup, PEAC allows one to estimate the differential phase $\theta$, and thus $a_\ext$.
We emphasise that phase information cannot be extracted from a single histogram when the intrinsic PDF parameters are not independently known; however, by considering multiple times $T$, the collapse of the amplitude $\amp_\all$ enables the extraction of the phase.
For comparison, as we have direct access to $\lambda_\mf$ from SG population measurements, we directly obtain $\lambda_0=\qty{0.39 (6)}{}$, $\lambda_{-1}=\qty{0.21 (3)}{}$, $\lambda_{+1}=\qty{0.40 (4)}{}$, $\Delta \lambda =0.19(5)$, and $\amp_0=\qty{0.83 (1)}{}$, averaged over all $T$ values.
These measured populations vary with every experimental realisation and are unknown, if no state-selective detection is performed.
While there are drifts on longer timescales, the population mixture appears constant within experimental uncertainties on the timescale of the performed experiments.
Here, $\amp_0$ is the average amplitude of the $\mf$ substates obtained by fitting the individual $\mf$ histograms.
These results are in good agreement with those obtained using PEAC.

If the populations $\lambda_\mf$ and amplitude $\amp_0$ are known, \eg{} from other, independent calibration measurements or the state preparation procedure, the differential phase can be estimated from a single set of data for a fixed interferometer time $T$, so that no control over the differential phase is necessary to implement PEAC.
With these quantities at hand, Eq.~\eqref{eq:A_all} can be applied pointwise, so that we demonstrate PEAC in static mode in Sect.~\ref{sec:correlation}.
Last, we highlight that PEAC can even be applied to systems with more than three states, albeit at the cost of a richer beat structure of the amplitude $\amp_\all$ (\cf{} \mm).

\subsubsection{Velocimetry}
\label{subsubsec:velocimetry}
\begin{figure}[!h]
    \centering
    \includegraphics[scale=.9]{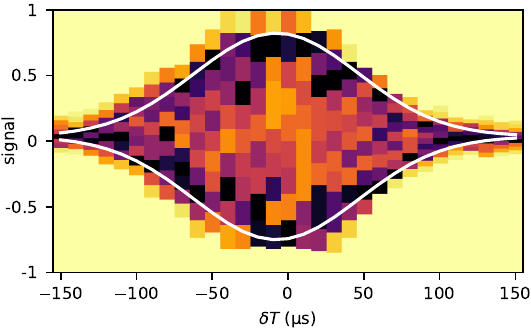}
    \caption{\textbf{Amplitude collapse in an open interferometer.}
    Density plot of an open interferometer signal for the $\mf = 0$ component.
    Increasing timing asymmetry $|\delta T|$ reduces the overlap of the two interferometer arms in the output, leading to an amplitude collapse.
    The white solid line traces the envelope of the amplitude collapse used to extract the momentum width of the atomic cloud.
        }
    \label{fig:2}
\end{figure}

In the previous section the observed amplitude collapse was caused by a beat signal of three correlated interferometer signals encoding a differential phase.
We now demonstrate the broad applicability of PEAC by performing velocimetry of atomic clouds to determine their momentum width $\Delta p$.
Since we expect the same behaviour for all $\mf$ substates, we restrict the analysis to the $\mf=0$ component extracted from a state-selective detection via the SG method.
Similar to contrast envelope fitting~\cite{mueller:2008:atom, kovachy:2015:quantum, castanet:2024:atom}, the interferometer is deliberately opened~\cite{roura:2014:overcoming} according to Fig.~\ref{fig:1}(\textbf{A}) by shifting the last $\pi/2$ pulse by $\delta T\in\left[\qty{-150}{\textmu s},\qty{150}{\textmu s}\right]$ with a step size of $\qty{10}{\textmu s}$.
We operate the interferometer at a fixed value of $T=\qty{3}{ms}$ and record five phase scans as described in Sect.~\ref{subsubsec:magnetometry}, totalling to $100$ realisations per $\delta T$.
The timing asymmetry reduces the spatial overlap of the wave packets propagating along the two arms at the output ports.
The resulting displacement scales as $ \delta \zeta = \hbar k_\eff \delta T/m$ with the mass $m$ of \rb{} atoms.
Assuming a Gaussian momentum distribution with width $\Delta p$, the amplitude of the interference signal $\amp(\delta T) = \amp_0 \exp \bigl[- \delta \zeta^2 \Delta p^2/(2\hbar^2) \bigr] = \amp_0 \exp \bigl[- \delta T^2/(2 T_\text{c}^2) \bigr] $ is suppressed by a Gaussian envelope with an apparent coherence time $T_\text{c} = m / (k_\eff \Delta p)$.
Similar to magnetometry, we apply PEAC by extracting the amplitude $\amp(\delta T)$ from fits of the PDF to the histograms of the interferometer signal, again employing thousandfold bootstrapping.
These amplitudes are then fitted to the Gaussian envelope to extract $T_\text{c}$.
In Fig.~\ref{fig:2} we show the density plots of the resulting histograms with white solid lines tracing the envelope $\amp(\delta T)$ for the averaged fit parameters.
We obtain $\amp_0=0.79(2)$ and $T_\text{c}=\qty{54.4(15)}{\textmu s}$, with the centre of the Gaussian shifted by $\qty{7.6(1.5)}{\textmu s}$.
Given this apparent coherence time, we estimate the rms momentum spread as $\Delta p = \qty{0.100(3)}{\ensuremath{\hbar k_\eff}}$, which translates to an effective temperature of $\Delta p^2 / (m k_\text{B})=\qty{13.7(7)}{nK}$.
These results agree with the widths obtained by TOF measurements ($\Delta p = \qty{0.12(1)}{\ensuremath{\hbar k_\eff}}$), with the remaining difference attributed to a broadening of the ensemble by resonant absorption imaging and contributions of the thermal component \cite{turpin:2015:bluedetuned}.
Using the timing asymmetry $\delta T$ as a control parameter, PEAC therefore enables the determination of the momentum width of an atomic ensemble from the collapse of the amplitude in a deliberately opened interferometer.

\subsection{Correlation Measurements -- Connection to Ellipse Fitting}
\label{sec:correlation}
As demonstrated in Sect.~\ref{subsubsec:magnetometry}, PEAC can be used to extract differential phases from the overlay and resulting beat of correlated signals in the scanning mode.
Here, we turn to the static mode establishing a connection to ellipse fitting.
In addition to emphasising the application to systems without state-selective measurement capabilities, we now shift our focus to correlated systems in a setup that allows for state selection, where all intrinsic parameters of the PDF have been obtained from independent measurements.
In this case, no control over the quantity of interest, \ie{} the differential phase $\theta$, is required, and PEAC can be applied to each data set at fixed interrogation time $T$.
Acknowledging that correlation measurements also can be carried out in systems with more than two distinct components~\cite{rosi:2015:measurement,xiaoyi:2005:fitting,rosi:2017:quantum}, we restrict our evaluation to a two-state system, similar to typical quantum clocks using $\pi/2$ pulses to generate coherent superpositions between two internal states.
We immediately can investigate this in our system by applying the SG method to distinguish the signals $S_\mf$ and choosing a symmetric sum of the $\mf= \pm 1$ states while excluding the magnetically insensitive $\mf=0$ state ($\lambda_0=0$).
The summed signal
\begin{equation}
    S_\summ = (S_{-1} + S_{+1})/\sqrt{2},
\end{equation}
is scaled by choosing $\lambda_{\pm 1}=1/\sqrt{2}$, anticipating Eq.~\eqref{eq:S_alpha}.
Evaluating Eq.~\eqref{eq:A_all} for this system and $\delta T =0$ gives the amplitude modulation~\cite{loriani:2019:interference,zych:2011:quantum}
\begin{equation}
\label{eq:A_sum}
    \amp_\summ= \sqrt{2}\amp_0 |\cos(\theta/2)|.
\end{equation}

\begin{figure}[h]
    \centering
    \includegraphics[scale=.9]{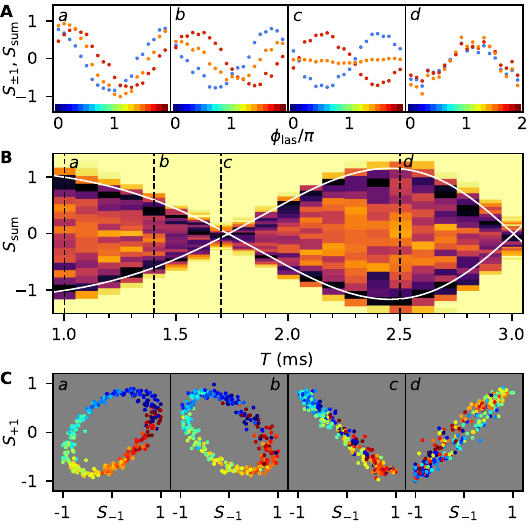}
    \caption{
    \textbf{Interference fringes, amplitude collapse, and signal correlation. }%
    (\textbf{A}) Coloured dots in panels (a) to (d) show the averaged interference fringes of $S_{+1}$ (red), $S_{-1}$ (blue), and $S_\summ$ (orange) for four different values of $T$, \ie{} for distinct differential phases $\theta$.
    The beat of $S_{+1}$ and $S_{-1}$ leads to a collapsing amplitude of $S_\summ$ for $\theta\cong\pi$ (c).
    Uncertainties are omitted for visual clarity.
    (\textbf{B})
    Collapse and revival of the amplitude $\amp_\summ$ (white) as a function of $T$, inferred from the underlying histograms of $S_\summ$ shown as density plot, where darker regions correspond to higher bin counts.
    The collapses occur at the beat nodes around $T=\qty{1.7}{ms}$ and $T=\qty{3}{ms}$.
    Dashed black lines indicate the times (a) to (d) from (\textbf{A}).
    (\textbf{C}) Bivariate scatter plots for times (a) to (d) of the correlated signals $S_{+1}$ (vertical) and $S_{-1}$ (horizontal), with a colour coding as in (\textbf{A}) indicating the phase scan of $\phi_\laser$.
    The eccentricity of the emerging ellipse encodes the differential phase $\theta$.
    At $T=\qty{1.7}{ms}$ $(\theta\cong\pi)$ and $T=\qty{2.5}{ms}$ $(\theta \cong 2 \pi)$, the ellipses become degenerate and collapse to lines along the anti-diagonal and diagonal, respectively.
    }
    \label{fig:3}
\end{figure}

In panels (a) to (d) of Fig.~\ref{fig:3}(\textbf{A}), we plot the averaged fringes of the individual signals $S_{\pm 1}$ and $S_\summ$ for four distinct interferometer times $T \in \{\qty{1.0}{ms},\qty{1.4}{ms},\qty{1.7}{ms},\qty{2.5}{ms}\}$, \ie{} different phases $\theta$.
For increasing $T$, phase noise induces a reduction in the observed fringe amplitudes of $S_{+1}$ and $S_{-1}$, which are equal in magnitude.
The observed fringe amplitude of $S_\summ$ decreases accordingly, but is in addition modulated by $\lvert \cos(\theta/2) \rvert$ and depends on the phase difference.

The signals $S_{\pm1}$ range from being slightly (a) and notably (b) out-of-phase, via a vanishing $ S_\summ$ in (c) for a differential phase $\theta\cong\pi$, to a rephasing at $\theta \cong 2 \pi$ in (d).
Resorting again to PEAC, we show the corresponding density plot of the histograms in Fig.~\ref{fig:3}(\textbf{B}), with black dashed lines marking the chosen times from Fig.~\ref{fig:3}(\textbf{A}).
This visualisation shows clearly the collapse and revival of the amplitude $\amp_\summ$ with $T$ and by that with varying differential phase $\theta$.
The beat node of the signal at $T=\qty{1.7}{ms}$ ($\theta\cong\pi$) becomes visible, with a secondary node discernible at $T=\qty{3}{ms}$ ($\theta \cong 3 \pi$).
Fitting the histogram PDF to $S_\summ$ for each interferometer time $T$ returns the time-dependent amplitude $\amp_\summ(T)$, while fits to $S_{\pm1}$ can be used to obtain $\amp_0$.
The white line in Fig.~\ref{fig:3}(\textbf{B}) shows the envelope given by Eq.~\eqref{eq:A_sum} for averaged values of $\amp_0=0.82(1)$ and $a_\ext=\qty{32.2(1)}{mm\pss}$, obtained by bootstrapping the data and fitting.
This envelope corresponds to the intrinsic amplitude of the signal, which is larger than the observed fringe amplitude, as the analysis of histograms is robust against phase noise.

Beside fitting the amplitude modulation, a known $\amp_0$ allows the pointwise estimation of the differential phase $\theta (T)$ by inverting Eq.~\eqref{eq:A_sum}.
To account for the ambiguity of the inverted cosine function, we unwrap the phase.
Fitting Eq.~\eqref{eq:theta} to the reconstructed phases returns an external acceleration of $a_\ext=\qty{32.2(1)}{mm\pss}$, which is in excellent agreement with our prior results.

We compare PEAC to a more established evaluation scheme for correlated systems based on parametric plots and fitting ellipses~\cite{foster:2002:method, chiow:2009:ellipse, fitzgibbon:1999:direct, halir:1998:numerically, halir_flusser:2025:ellipse, ridley:2024:investigation, zhang:2023:dependence}.
The correlated signals $S_{\pm 1}$ for the interferometer times of Fig.~\ref{fig:3}(\textbf{A}) are visualised in a bivariate scatter plot in Fig.~\ref{fig:3}(\textbf{C}), with $S_{-1}$ along the horizontal and $S_{+1}$ along the vertical axis; each measurement realisation is shown as a coloured dot.
The colour is chosen to represent the adjusted phase $\phi_\laser$, which results in a colour gradient for the performed phase scan as can be seen in sub-figures (a) to (d).
For larger interferometer times $T>\qty{1.6}{ms}$, the colour gradient becomes disordered, showing a transition to the phase-unstable regime of the experiment.
While a scan of the phase $\phi_\laser$ traces the ellipse, the differential phase is encoded in its eccentricity, which we extract from geometric ellipse fitting routines~\cite{gander:1994:ellipses, ahn:2001:geom_ellipse, corgier:2025:optimized, meister:2025:space}.
A detailed explanation of our fitting routine can be found in \mm.
With this approach, we estimate an external acceleration of $a_\ext=\qty{32.12(2)}{mm\pss}$, confirming the results obtained by PEAC.

\begin{figure}[htb]
    \centering
    \includegraphics[scale=.9]{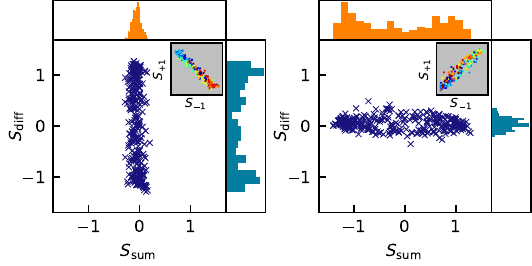}
    \caption{\textbf{Rotated bivariate scatter plots and histograms.}
    Rotating the original correlation signals (insets) for $T=\qty{1.7}{ms}$ ($\theta\cong\pi$, left) and $T=\qty{2.5}{ms}$ ($\theta \cong 2 \pi$, right) by $\pi/4$ clockwise aligns the ellipses to their principal axes, which correspond to $S_\summ$ (horizontal) and $S_\diff$ (vertical).
    The marginal distributions show the histograms of $S_\summ$ and $S_\diff$, illustrating that the two exchange their roles at the respective degeneracy points of $\theta$.
    This representation visualises the intrinsic link between ellipse-based estimation based on bivariate fitting and PEAC applied to $S_\summ$ and $S_\diff$.
    Note that $S_\diff$ is only accessible by state-selective detection.
    }
    \label{fig:4}
\end{figure}

In fact, both methods are inherently connected:
The principal axes of the ellipses in Fig~\ref{fig:3}(\textbf{C}) (c) and (d) are given by the diagonal and anti-diagonal, see \mm{} for more details.
The diagonal is proportional to the sum of the bivariate coordinates $(S_{+1}+S_{-1})$ while the anti-diagonal is proportional to their difference $(S_{+1}-S_{-1})$.
Rotating the bivariate coordinate system by the angle $\alpha$ leads to the definition of the general signal
\begin{equation}
\label{eq:S_alpha}
    S(\alpha) = S_{-1}\cos\alpha +  S_{+1}\sin\alpha.
\end{equation}
Hence, the special case $S_\summ= S(\pi/4)$ denotes the axis along the diagonal and can be accessed even without state-selective measurements for two-component systems.
A second case is $S_\diff= S(3\pi/4)$, oriented along the anti-diagonal, which can only be inferred in a state-selective setup.
Consequently, for two correlated states such as presented here, applying PEAC generalises to performing a statistical histogram analysis of the bivariate data along any angle $\alpha$.
In Fig.~\ref{fig:4} we depict the ellipses aligned to their principal axes $S_\summ$ and $S_\diff$, \ie{} rotated by $\pi/4$ clockwise, for $T=\qty{1.7}{ms}$ (left) and $T=\qty{2.5}{ms}$ (right) in the bivariate variables $(S_\summ, S_\diff)$ with complementing histograms at the marginals.
The displayed ellipses are close to degeneracy, \ie{} $\theta\cong\pi$ (left) and $\theta\cong2\pi$ (right), where the method of ellipse fitting has known limitations.
Especially large baseline fluctuations lead to large biases of the estimated phase close to these degeneracy points.
Generally, the rotation angle $\alpha$ encodes the population ratio of the two involved states and reflects any imbalance due to imperfect state preparation.
In addition, for any recorded bivariate data set, histograms can be constructed and analysed along a chosen direction $\alpha$, complementing the ellipse-based analysis.
For a detailed analysis of different choices of $\alpha$, we refer to the \supp, but report here the key insights.
From simple Gaussian uncertainty propagation, one expects that $S_\summ$ is the optimal choice.
However, this idea neglects a possible $\alpha$-dependence of the amplitude uncertainty arising from the explicit fitting procedure.
Taking this into account shows that PEAC can benefit from a slight deviation from $\alpha=\pi/4$ in terms of a better trueness over a wider range.
The optimal choice of $\alpha$ can generally depend on the specific use case and application, such that we refrain to specify a single best value here and focus on $\alpha=\pi/4$ for the remainder.
In the following section, we investigate the systematic error and compare PEAC in trueness and precision to the ellipse fitting approach.

\subsection{PEAC Performance}
\label{sec:PEAC_perf}

\begin{figure*}[htb]
    \centering
    \includegraphics[scale=0.90]{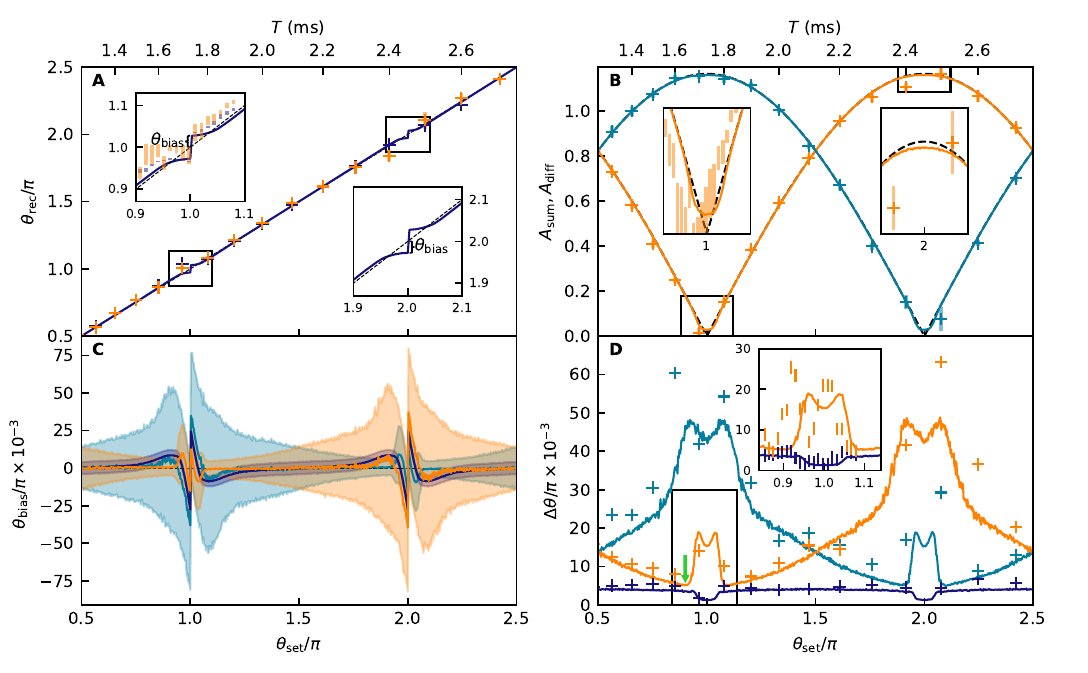}
    \caption{\textbf{Performance of phase estimation: Ellipse fitting versus PEAC.}
        Comparison of ellipse fitting (dark blue) and PEAC along the $S_\summ$ (orange) and $S_\diff$ (light blue) directions.
        Experimental data are shown as crosses (uncertainty bars omitted for clarity) or rectangular bars, with centres indicating the mean and heights equal to twice the bootstrapped standard deviation.
        Solid lines are values obtained from numerical replications of the experiment.
        The dashed lines correspond to the set phases and amplitudes as reference values free of fitting errors.
        We convert the set phase $\theta_\set$ via Eq.~\eqref{eq:theta} to $T$.
        (\textbf{A}) Reconstructed phase.
        The inset shows that ellipse fitting leads to a large bias $\theta_\bias = \theta_\recons-\theta_\set$ near degeneracy, despite high precision, while PEAC along the minor axis $S_\summ$ exhibits larger uncertainty but a reduced bias.
        (\textbf{B}) Modulation of the amplitudes $A_\summ$ and $A_\diff$  obtained from histogram fits along the two principal axes.
        We observe a deviation at degeneracy, contributing to PEAC's remaining bias.
        (\textbf{C}) Bias obtained from numerical replications.
        PEAC along the favourable (minor) axis increases the effective bias-free region by a factor of $1.6$ compared to ellipse fits, substantially enhancing trueness.
        The semi-transparent bands give the uncertainties extracted from $1000$ replicated experiments.
        (\textbf{D}) Precision of phase estimation.
        Ellipse fitting maintains low uncertainty across all $\theta_\set$, while PEAC of $S_\summ$ achieves comparable precision only near, but not at the degeneracy point $\theta_\set = \pi$, as would be expected from simple Gaussian uncertainty propagation.
        This behaviour arises from an ambiguity in the histogram's PDF for small amplitudes.
        For PEAC, a distinct minimal uncertainty can be found, marking an optimal working point (see green arrow).
        The visible reduction in the ellipse's uncertainty around the degeneracy points stems from the least-squares fitting routine converging to the boundary condition for the major semi axis.
        }
    \label{fig:5}
\end{figure*}

To quantify the performance of PEAC, we assess the trueness and precision of phase estimation, without relying on scanning the differential phase.
Figure~\ref{fig:5}(\textbf{A}) displays the unwrapped phase $\theta_\recons$, reconstructed individually from the experimental data for each $T$ from geometric ellipse fitting (dark blue crosses, masked by orange crosses for most data points), see {\mm} for details.
For a complementary theoretical analysis establishing a connection between experiment and theory, we use the control variable $\theta_\set$, which we convert to the corresponding $T$ via Eq.~\eqref{eq:theta} using $a_\ext=\qty{32.2(1)}{mm\pss}$.
Ideally, we expect the black dashed line representing identity, corresponding to perfect phase reconstruction $\theta_\recons = \theta_\set$, yet notable deviations of the experimental data occur.
This discrepancy becomes more evident for the numerical replication of the experiment, shown as the dark blue line.
We use the experimentally inferred parameters characterising the signals (see details in \mm).
In essence, the numerical replication allows for a much finer sampling of $\theta_\set$.
The largest discrepancies occur at $\theta_\set = \pi \mathds{Z}$, \ie{} at points of degeneracy, which are a fundamental limitation to ellipse-based phase estimation~\cite{foster:2002:method,ridley:2024:investigation}.
We experimentally study one of these degeneracy points in more detail: The inset at $\theta_\set=\pi$ illustrates this discrepancy through a finer experimental sampling in $T$.
In this inset, the centre of each rectangle denotes the mean, while its height is twice the empirical standard deviation extracted from bootstrapping, caused by the statistical nature of the baseline fluctuations.
The experimental data obtained from ellipse fitting (dark blue) exhibits the same qualitative behaviour as the numerical replication, thereby confirming the validity of our theoretical model.
Hence, a bias $\theta_\bias = \theta_\recons-\theta_\set$ in the vicinity of the degeneracy points deteriorates the trueness of the estimated phase.

Figure~\ref{fig:5}(\textbf{B}) presents the amplitudes $\amp_\summ$ (orange) and $\amp_\diff$ (blue) of $S_\summ$ and $S_\diff$, extracted from histogram fits.
The data clearly underline the collapse and revival behaviour.
Fits to the  numerical replication lead to the solid lines for $\amp_\summ$ (orange) and $\amp_\diff$ (blue), which are phase shifted by $\pi$, as expected from Fig.~\ref{fig:4}.
The insets magnify the deviation near the degeneracy points.
As in panel (\textbf{A}), we present the ideal amplitudes $\amp_\summ= \sqrt{2}\amp_0 |\cos(\theta_\set/2)|$ and $\amp_\diff= \sqrt{2}\amp_0 |\sin(\theta_{\set}/2)|$ by black dashed lines.
Applying PEAC to the experimental data, we infer $\theta_\recons(T)$ via Eq.~\eqref{eq:A_sum} and thus obtain the orange crosses in panel (\textbf{A}) and rectangles in its inset.
See \mm{} for a detailed description of the full protocol.
Compared to the reconstruction via the ellipse method, we see a larger standard deviation, and by that a decrease in precision.

The numerical replication allows us to quantify the respective bias $\theta_\bias$ of the different estimation methods, shown in Fig.~\ref{fig:5}(\textbf{C}) as solid lines (dark blue: ellipse, orange: $S_\summ$, blue: $S_\diff$).
Enabled by state-selective detection through the SG method and post-processing of the data, the difference signal $S_\diff$ is also analysed using PEAC, revealing that the bias is $\pi$-phase shifted relative to $S_\summ$.
The traces directly show the averaged bias and are a measure of trueness (high if $\theta_\bias \cong 0$), while the shaded areas represent the uncertainty $\pm \Delta \theta$ in phase reconstruction, obtained from statistical repetitions as detailed in \mm{}, and quantify the precision: smaller uncertainty bands correspond to higher precision.
This plot illustrates the challenges of the ellipse fitting method:
While it can achieve high precision, it suffers from major bias leading to low trueness, particularly at points of degeneracy.
In fact, for our data, the ellipse fitting method has negligible bias except for an interval of  $\pm\pi/4$ around the degeneracy points.
In contrast, PEAC substantially increases the bias-free region, so that a bias is only observed in an interval $\pm\pi/10$ around degeneracy and therefore enhances the trueness at the expense of precision.
We also observe that $S_\summ$ and $S_\diff$ reverse their roles at opposing degeneracy points, such that applying PEAC to $S_\summ$ around $\theta \cong \pi$ benefits from higher precision, while analysing  $S_\diff$ is favourable around $\theta \cong 2\pi$.
Considering the bivariate $(S_\summ, S_\diff)$ plane of Fig.~\ref{fig:4}, we understand this behaviour geometrically:
The bias and precision are only substantially reduced when the histogram is obtained by projecting along the minor axis of the ellipse.
We give more insights into the bias's origin in \mm.
Along the favourable minor axis, PEAC extends the range of phases for which effectively bias-free estimation is possible by a factor of $1.6$ compared to ellipse fitting.
Even along the major axis, PEAC performs slightly better than ellipse fitting, although with a high uncertainty, which depends on the number of measuring points and further optimisation schemes \cite{pelluet:2025:atom}.
For phases within $\pm \pi/10$ around degeneracy we observe considerable variations in bias across all three cases.
Since the variation amplitude is the smallest for the favourable direction of PEAC, it maintains a reduced bias and remains the optimal choice for phase estimation in this regime.
Thus, PEAC achieves higher trueness compared to the ellipse method, which is a necessary condition for high-accuracy measurements, for which both trueness and precision are essential.

Last, we compare the precision of both methods by investigating the phase uncertainty $\Delta \theta$ shown as shaded areas in panel (\textbf{C}).
The values of the uncertainty are displayed as solid lines in Fig.~\ref{fig:5}(\textbf{D}) for a direct comparison.
To benchmark the theoretical analysis, we also add the corresponding experimental values as crosses and vertical bars in the inset around degeneracy (dark blue: ellipse fitting, orange: $S_\summ$, blue: $S_\diff$).
The uncertainty of the ellipse method is almost constant across all phases and consistently lower than that of PEAC, for both $S_\summ$ and $S_\diff$.
The dip of its uncertainty around degeneracy arises because the geometric ellipse fits are constrained by boundaries that enforce physically reasonable, \ie{} normalised, principal axes.
Apparently, the tendency to converge to these boundaries reduces the statistical uncertainty.
Nevertheless, PEAC can achieve uncertainty levels as low as ellipse fitting for specific phases near, but not exactly at degeneracy.
In quantum clock interferometry, simple Gaussian uncertainty propagation suggests $\Delta \theta = \Delta \amp_\summ / |\partial_\theta \amp_\summ|$, where  $\Delta \amp_\summ$ is the uncertainty of the oscillation amplitude.
Following this intuition, the amplitude $\amp_\summ= \sqrt{2}\amp_0 |\cos(\theta/2)|$ exhibits its steepest slope and by that the lowest uncertainty at $\theta=\pi$.
In contrast, a vanishing slope at $\theta=2\pi$ implies a diverging phase uncertainty and explains why the major axis is unfavourable for PEAC along $S_\summ$.
However, as shown in the inset of Fig.~\ref{fig:5}(\textbf{B}), we observe that near the degeneracy point $\theta_\set \cong \pi$, the fitted value of $\amp_\summ$ deviates from the expected behaviour.
This deviation and the connected increase in uncertainty arise from a merging of the two maxima of the PDF for low amplitudes, resulting in a single-peak PDF.
Although no closed analytical expression describes this transition, a numerical analysis reveals that it occurs when $\amp/\sigma < 1.78$, where $\sigma$ is the standard deviation of the baseline fluctuations.
Since $A$ depends on $\theta$, this transition from a double-peaked to a single-peaked PDF can be associated with a differential phase, which for our experimental parameters is $0.94 \pi$.
This differential phase aligns well with our observation of an increase in both bias and uncertainty.
To show that this observation holds for various settings, we performed a scan of the baseline noise $\sigma$.
The corresponding two-dimensional scans are summarised and compared against ellipses in the \supp.
Indeed, we observe that both bias and uncertainty of the phase estimation based on PEAC depend on baseline fluctuations.
While ellipses show a superior behaviour for vanishing noise, PEAC outperforms ellipse fits at higher noise levels.
We observe that substantial bias is confined to a narrow region around degeneracy, where $A < 1.78 \sigma$, and increases linearly with $\sigma$.
In this region, the increased ambiguity in the fit parameters $\amp_\summ$ and $\sigma$ results in larger bias $\theta_\mathrm{bias}$ and increased phase uncertainty $\Delta \theta$.
Therefore, despite the expectation from Gaussian uncertainty propagation, the minimal phase uncertainty is not observed at the degeneracy point $\theta_\set = \pi$, \ie{} at vanishing signal amplitude.
In fact, minimal uncertainty is reached near, but not at the aforementioned degeneracy point, namely where the signal amplitude is small but still finite.
These insights allow us to define the working point to be that phase with no substantial bias and lowest phase uncertainty (see Fig.~\ref{fig:5}(\textbf{D}) green arrow).
From our baseline-noise scan we infer that the working point is in close proximity to the transition from a single-peak to a double-peak PDF.
At this point, the level of uncertainty is similar to that of ellipse fitting, but the trueness is superior for non-vanishing baseline noise.
Our analysis in the \supp{} further indicates that this working point exhibits a linear scaling with $\sigma$.
In addition and to establish a robust criterion for when an estimate can be trusted, we also normalise the bias in units of uncertainty in the \supp, demonstrating that PEAC remains reliable throughout the entire interval in contrast to ellipses.

Naturally, these results depend not only on the specific noise and noise model but also on the chosen method of parameter estimation from a given PDF.
In this article, we use a standard least-squares-based fitting routine because of its simplicity and numerical performance.
However, other methods such as maximum likelihood estimation can provide better results.
To compare these two approaches, we perform an analysis of bias and uncertainty in analogy to Fig.~\ref{fig:5}(\textbf{C}) and (\textbf{D}) based on maximum likelihood estimation in the \supp.
In summary, we observe that this approach reduces both bias and uncertainty even further, albeit at the cost of a substantial increase in computational time.
Given this trade-off, we have utilised least-squares-based fitting for our primary evaluation.
However, the \supp{} demonstrates that PEAC is inherently compatible with more sophisticated inference techniques and therefore compatible with further future improvements.

\section{DISCUSSION}
Although the statistical analysis of the noisy signal of an interferometer, such as histogram-based estimation of the underlying probability density function (PDF), provides access to the oscillation amplitude and baseline fluctuations, it inherently lacks phase information.
When two such signals are correlated, common-mode noise is suppressed, allowing for the extraction of a differential phase $\theta$ from the bivariate data, for example from fitting to ellipses~\cite{foster:2002:method, chiow:2009:ellipse, fitzgibbon:1999:direct, halir:1998:numerically, halir_flusser:2025:ellipse, ridley:2024:investigation, zhang:2023:dependence} or Lissajous curves~\cite{barret:2015:correlative,chen:2014:lissajous}.

In this work, we establish a geometric connection between these two paradigms:
We introduce Parameter Estimation from Amplitude Collapse (PEAC), a method that performs a statistical analysis along a suitably chosen projection direction in the bivariate plane.
It extracts a differential phase from the observed amplitude, which shows a characteristic collapse and revival due to the beat of overlaying signals.
Unlike conventional ellipse fitting, which becomes ill-posed near degeneracy points ($\theta = \pi \mathds{Z}$) where the ellipse collapses to a noisy line~\cite{ridley:2024:investigation}, PEAC always applies a statistical inference technique to the marginal distribution.
This procedure eliminates the issues of a geometric degeneracy and reduces systematic bias, thereby enhancing trueness and overall accuracy.
We highlight five key advancements:

\setlength{\parindent}{12pt}
(i) PEAC allows extracting differential phases and amplitudes in setups where the individual signals cannot be resolved independently, for example in non-state-selective setups where the individual components overlap spatially, overcoming one limitation of conventional ellipse fitting.
Importantly, while we demonstrated PEAC by scanning the differential phase in Sect.~\ref{sec:PEAC} to characterise the signal, such a scan is not a mandatory requirement.
Provided that the system's relevant parameters, such as the weights $\lambda_\mf$ and the intrinsic amplitude $A_0$, are determined through independent measurements or initial calibrations, the differential phase can be inferred without an explicit scan.
This flexibility is particularly advantageous in tests of fundamental physics or quantum clock interferometry, where the measurand may be minute and modulation of the differential phase is often experimentally impractical.

(ii) It is a complementary analysis applicable to any system employing ellipse fitting without the need for additional experimental resources.
While it generally requires more data points than ellipse fitting due to its statistical nature, PEAC becomes a powerful tool when sufficient statistics are available.
It generalises naturally by projecting along an angle $\alpha$ in the bivariate plane, enabling phase estimation of arbitrary mixtures of the correlated signals.
In fact, for a given set of bivariate data, one can always perform PEAC along a different angle in addition to ellipse fitting (see Sect.~\ref{sec:correlation} in combination with the \supp) to improve the trueness of the estimated phase by reducing the bias and systematic error near degeneracy.
While ellipse-based estimation is typically only unbiased at circular degeneracy points ($\theta = \pi/2$ and odd multiples), PEAC provides a vanishing bias throughout the $2\pi$ interval of $\theta$, except for a region of $A < 1.78 \sigma$ around degeneracy.
Consequently, PEAC is beneficial when the differential phase cannot be precisely controlled.
Furthermore, in Sect. \ref{sec:correlation}, all parameters of the PDF were determined from independent measurements, experimentally demonstrating that differential phase estimation is possible without the necessity of controlling or scanning the measurand.
A reduced bias closer to degeneracy is not merely a technical improvement, but critical for high-accuracy null measurements, for example envisioned for tests of fundamental physics~\cite{rosi:2014:precision,lamporesi:2008:determination,fixler:2007:atom, pikovski:2015:uni_decoh}.

(iii) We demonstrate that PEAC directly connects to quantum clock interferometry~\cite{roura:2020:gravitational,dipumpo:2021:gravitational,rosi:2017:quantum,fromonteil:2025:nonlocal, borregaard:2025:qclock, guendogan:2026:qclock} and answers the question of the optimal working point, which in fact is close to, but not exactly at degeneracy.
Notably, we observe no divergence of phase sensitivity close to degeneracy, as predicted from geometric phase amplification~\cite{zhou:2025:geometric} when neglecting the vanishing oscillation amplitude or from Gaussian uncertainty propagation.
Even in the context of fundamental physics applications, where one typically probes null signals or minute phase shifts and therefore has limited flexibility to access regimes with substantial amplitude collapse, PEAC has practical implications, as outlined above.
Furthermore, recent quantum clock experiments employing magnetic field gradients to measure geometric phases~\cite{zhou:2025:geometric} have observed a considerable reduction in contrast, highlighting the relevance of such methods in realistic scenarios.

(iv) Beyond two correlated signals, we demonstrate the applicability of PEAC to three correlated signals, without requiring to resolve them individually, by the example of a mixture of three magnetic sublevels.
The technique is therefore established as a complementary analysis of three- or higher-dimensional ellipse methods~\cite{rosi:2015:measurement} by analysing histograms along the semi axes.
Moreover, it can be applied to scenarios where an amplitude modulation is introduced deliberately into only one signal, situations that are inaccessible to ellipse fitting.
For example, we extend contrast-envelope fits~\cite{mueller:2008:atom, kovachy:2015:quantum, castanet:2024:atom}, a standard technique for atom interferometry, where the interferometer is partially or dynamically opened to control the visibility of the interference pattern.
Similarly, it applies to schemes where a phase modulation translates into a characteristic amplitude modulation~\cite{otabe:2025:sensitivity}.

(v) Last, PEAC is broadly applicable to a wide range of setups, from fully phase-stable to noisy, phase-unstable regimes.
The sensitivity of phase estimation can be enhanced by resorting to higher-order atom diffraction, demonstrating that it scales with increasing momentum transfer, and can be applied to periodic signals beyond atom interferometry, such as atomic clocks~\cite{zheng:2023:labbased, bothwell:2022:resolving} or optical interferometers.
Moreover, the statistical nature of the method offers potential for integration into hybrid classical-quantum sensors~\cite{templier:2022:tracking} or for locking devices to a regime of minimal phase uncertainty~\cite{chiow:2016:noise}.

We identified a working point for PEAC at phases close to, but not exactly at, degeneracy.
Using a numerical model and associated parameter scans, we showed that its exact location depends approximately linearly on the strength of baseline fluctuations.
However, although a preferred phase of operation exists, PEAC remains applicable even when the differential phase cannot be sufficiently controlled.
In this case, ellipse fitting exhibits a substantial bias across all phases except at $\theta = \pi/2$, exceeding the statistical uncertainty, whereas PEAC yields a bias-free estimate except in a narrow region around degeneracy.
By contrast, in the idealised absence of baseline fluctuations, ellipse fitting outperforms PEAC, as demonstrated by our numerical analysis.

Moving beyond the least-squares fitting of PDFs used in the main text, PEAC can be further improved by combining it with alternative statistical inference techniques, such as maximum likelihood estimation~\cite{pezze:2025:joint} discussed in the \supp.
Furthermore, it can be integrated with other approaches like Bayesian methods~\cite{stockton:2007:bayesian, pelluet:2025:atom, barret:2015:correlative}.
As these techniques pose a more targeted approach to parameter estimation, we anticipate a further reduction of the bias.
Similarly, PEAC can be combined with recent developments in ellipse fitting, such as geometric procedures that minimise the bias with the help of optimised cost functions~\cite{meister:2025:space}.
Using PEAC in combination with ellipse fitting and inducing a feedback loop between these procedures can help to decrease the bias in a joint phase estimation protocol, especially close to degeneracy.
One could further allow for adaptive bin sizes within a single histogram and automatically assess the number of required bins to resolve the important features of the PDF without increasing the statistics.
Such a procedure would also lead to improved initial guesses for the fitting routines.
PEAC is a classical statistical method, and the presented experimental results are obtained in a setup limited by technical noise one order of magnitude above shot noise.
Further reducing technical noise sources leads to a regime in which the sensitivity is limited by quantum projection noise, \ie{} shot noise.
In this case, the signal is no longer homoscedastic, and PEAC must be adapted accordingly.
At the same time, this regime constitutes the prerequisite for quantum-enhanced metrology.
In fact, combining PEAC with quantum input states, such as optimally squeezed states~\cite{corgier:2025:optimized}, has the potential to surpass current limits in quantum metrology and to advance perspectives in a broad range of applications.

\section{MATERIALS AND METHODS}
\subsection{Experimental System and Measurements}
Our experimental setup produces BECs in a crossed optical dipole trap (CDT) by forced evaporative cooling~\cite{lauber:2011:pra,pfeiffer:2025:novel}.
To perform measurements in free space, the atoms are released from the CDT into free fall.
Prior to the start of the MZI sequence, an expansion time of $\qty{3}{ms}$ allows for the conversion of the mean-field to kinetic energy.
After this expansion, typically~\cite{pfeiffer:2025:dichroic} the momentum spread of the ensemble is $\Delta p=\qty{0.13(3)}{\ensuremath{\hbar k_\eff}}$.
The laser beams used for Bragg diffraction are superimposed onto one of the CDT beams in a counter-propagating setup, oriented perpendicular to the direction of gravity.
This ensures an initial momentum of $p_0=0$ in the direction of Bragg diffraction.
To drive two-photon transitions, the Bragg beams are detuned by $\Delta=2\pi\times\qty{3.2}{GHz}$ from the excited state manifold $\ket{e} = \ket{5^2P_{3/2}, F=2}$, transferring momentum $\hbar k_\eff=\,2\hbar  \times 2\pi / \qty{780.226}{nm}$ along the Bragg axis.
For the results presented in this article, we rely on first-order Bragg diffraction by choosing a detuning $\Delta\omega=2\pi\times\qty{15.084}{kHz}$ between the two beams, while our setup also allows for higher-order diffraction \cite{pfeiffer:2025:dichroic} by adjusting $\Delta \omega$ to other resonances.
A Blackman window function $f(t) = 0.42 - 0.5\cos(2\pi t/\tau) + 0.08\cos(4\pi t/\tau)$ for $0\leq t\leq \tau$ as a smooth pulse shape ensures efficient transfer between the coupled momentum states and reduces the population of parasitic momenta.
We use a pulse length of $\tau=\qty{100}{\textmu s}$ as a compromise between velocity selectivity and diffraction efficiency, adjusting the peak Rabi frequency of the three pulses to adapt for the $\pi/2$--$\pi$--$\pi/2$ pulse configuration.
Following each MZI, a time of flight (TOF) of typically $\qty{15}{ms}$ to $\qty{30}{ms}$ is applied to separate the two interferometer ports spatially for detection.
We perform resonant absorption imaging~\cite{ketterle:1999:imaging} of the atomic densities to gain access to the atom numbers $N_{0}$ and $N_{1}$ in each of the interferometer ports.
Without optical pumping or a state-selective SG measurement, the densities of the three $\mf$ substates will spatially overlap and cannot be distinguished.
However, our setup also allows using the SG method, by applying a magnetic field gradient after the MZI, generated by one of the magnetic coils used for the magneto-optical trap, separating the $\mf$ substates spatially via the Zeeman force.
A time delay of $\qty{1}{ms}$ between the second $\pi/2$ pulse and the introduction of the magnetic field gradient prevents spurious phase contributions to the interferometer phase.
The applied weak gradient is left on for a total of $\qty{24}{ms}$ until detection to ensure proper state separation.

\subsection{Derivation of Signal $S$}
The normalised difference between the atom numbers $N_j(\phi) = \base_j \pm \amp_j \cos\phi$ with $j\in\{0,1\}$ detected in the two momentum classes leads to the signal
\begin{equation}
    S = \frac{N_0 - N_1}{N_0 + N_1} = \frac{\Delta \base + 2 \bar{\amp} \cos \phi}{2 \bar{\base} + \Delta\amp \cos \phi},
\end{equation}
introducing the mean and differential baseline $\bar{\base} = (\base_0 + \base_1)/2$ and $\Delta \base = \base_0 - \base_1$, as well as the mean and differential amplitude $\bar{\amp} = (\amp_0 + \amp_1)/2$ and $\Delta \amp = \amp_0 - \amp_1$, respectively.
With only two exit ports, the total atom number $N_0 + N_1$ must be preserved and independent of the interferometric phase $\phi$, implying $\amp_1 = \amp_0$.
As a result, we find
\begin{equation}
    S = \frac{\Delta \base}{2 \bar{\base}} + \frac{\amp_0}{\bar{\base}} \cos \phi.
\end{equation}
In general, $\amp_j, \base_j$ are suitable random variables describing experimental stochastic fluctuations of baselines and amplitudes, which may cause a heteroscedastic signal.

It is typically assumed that the baseline noise $\Delta \base/(2 \bar{\base})$ is normally distributed and dominant, such that $\amp_0/\bar{\base}$ can be treated as constant~\cite{pelluet:2025:atom}.
Observing a homoscedastic signal, \ie{} in our setting treating $\amp_0/\bar{\base}$ as constant, is possible if both random variables are correlated.
Assuming that $\amp_0$ and $\bar{\base}$ are perfectly correlated gives
\begin{equation}
\label{eq:amp_baseline_corr}
    \amp_0 = \sqrt{\frac{\mathbb{V}[\amp_0]}{\mathbb{V}[\bar{\base}]}} \bigl( \bar{\base} - \mathbb{E}[\bar{\base}] \bigr) + \mathbb{E}[\amp_0],
\end{equation}
where $\mathbb{E}[\cdot]$ denotes the expectation value and $\mathbb{V}[\cdot]$ the variance of a random variable.
To have constant $\amp_0/\bar{\base}$ and excluding the trivial constant signal, \ie{} $\mathbb{E}[\amp_0] = 0$, the condition
\begin{equation}
\label{eq:mean_var_quot}
    \frac{\mathbb{E}[\amp_0]}{\mathbb{E}[\bar{\base}]} = \sqrt{\frac{\mathbb{V}[\amp_0]}{\mathbb{V}[\bar{\base}]}}
\end{equation}
must be satisfied.
With Eq.~\eqref{eq:mean_var_quot}, we observe from the condition of perfect correlation \eqref{eq:amp_baseline_corr} that the signal takes the form
\begin{equation}
    S = \frac{\Delta \base}{2 \bar{\base}} + \frac{\mathbb{E}[\amp_0]}{\mathbb{E}[\bar{\base}]} \cos \phi,
\end{equation}
in which only $\Delta \base/ (2 \bar{\base})$ fluctuates.

Assuming normally distributed baselines $\base_j \sim \mathcal{N}(\mu_j, \sigma_j^2)$ with mean $\mu_j$ and standard deviation $\sigma_j$, for resolving a signal, the fluctuations must be sufficiently small, \ie{} $\mu_j \gg \sigma_j$.
If we additionally assume the same baseline fluctuations in both exits of the interferometer, \ie{} $\sigma_1 = \sigma_2 $, we can treat the mean baseline as constant $\bar{\base} \sim \mathcal{N} \bigl(\bar{\mu}, \sigma_1^2/2 \bigr) \cong \bar{\mu} = \mathbb{E}[\bar{B}]$ with $\bar{\mu} = (\mu_1 + \mu_2)/2$ and $\sigma_1=\sigma_2\ll \bar \mu$.
With these stipulations we finally arrive at a homoscedastic signal
\begin{equation}
    S = \base + \amp \cos \phi,
\end{equation}
with $\base = \Delta \base/(2 \mathbb{E}[\bar{\base}]) \sim \mathcal{N}\bigl( \mu, \sigma^2 \bigr)$ of mean $\mu = (\mu_1-\mu_2)/(2\bar{\mu})$ and variance $\sigma^2 = \sigma_1^2/(2 \bar{\mu}^2)$, and $A = \mathbb{E}[\amp_0]/\mathbb{E}[\bar{\base}]$, that has no amplitude fluctuations and only normally distributed baseline fluctuations.

\subsection{Finite-Pulse Duration}
The finite duration $\tau$ of the Bragg pulses can be a substantial fraction of the interferometer time $T$, ranging from $\qty{3}{\%}$ up to $\qty{10}{\%}$ in our experiments, and therefore introduces a non-negligible phase contribution~\cite{bertoldi:2019:phase,antoine:2006:matter,Li:2015:raman, bott:2023:atomic} that must be included in our phase analysis.
To incorporate this phase contribution, we resort to a numerical treatment, necessary for time-dependent pulses, and define the accumulated time-dependent pulse area
\begin{equation}
\phi_1(t) = \int \displaylimits_{0}^t \! \dd t' \, \Omega(t')
\end{equation}
with the time-dependent Rabi frequency train $\Omega(t)= \Omega_0 [f(t)+ 2 f(t-T)+f(t-2T)]$ described by the Blackman envelope $f$.
As in the experiment, we choose $0.42\,\Omega_0\tau=\pi/2$ accounting for the $\pi/2$ pulses at $t=0$ and $t=2T$ as well as for the $\pi$ pulse at $t=T$.
An external acceleration $a_\eff$ induces a time-dependent Doppler detuning $k_\eff a_\eff t$, assuming the atoms are initially on resonance in agreement with our experimental realisation.
For example, to obtain the phase of the $\mf=+1$ state induced by $ a_\eff$ we evaluate~\cite{bertoldi:2019:phase}
\begin{equation}
\label{eq:theta_finite_pulse}
    \theta(T)/2 = \int \displaylimits_{0}^{2T+\tau} \! \dd t \, k_\eff a_\ext t \sin [\phi_1(t)]
\end{equation}
numerically for $\tau=\qty{100}{\textmu s}$ and various $T\in \left[\qty{1}{ms},\qty{3}{ms}\right]$.
For a simplified analytical expression, we fit $\theta(T) = 2 k_\eff a_\eff T^2(1+\gamma\tau/T)$ to the resulting values and obtain the fit parameter $\gamma \cong 0.1486$.

\subsection{General Amplitude Modulation}
Here, we present the derivation of the amplitude modulation $\amp_\all$ in Eq.~\eqref{eq:A_all} of the main text.
A sum of $n$ cosine functions weighted by $\lambda_i \in \mathds{R}$ with a common phase $\phi_0$, but possible different phases $\theta_i$, results~\cite{weisstein:harm_add_thm} in
\begin{subequations}
\label{eq:sum_of_cosines_all}
\begin{equation}
\label{eq:sum_of_cosines}
    \sum_{i=1}^n \lambda_i \cos(\phi_0 + \theta_i) = \amp[\{ \lambda_i \}, \{ \theta_i \}] \cos(\phi_0 + \theta_\offset) \\
\end{equation}
with the modulated amplitude
\begin{equation}
\label{eq:sum_of_cosines_ampl}
    \amp^2 = \sum_{i=1}^n \lambda_i^2 + 2 \sum_{i=1}^n \sum_{j > i}^n \lambda_i \lambda_j \cos(\theta_i - \theta_j)
\end{equation}
and offset phase
\begin{equation}
\label{eq:sum_of_cosines_offset_phase}
    \tan\theta_\offset = \frac{\sum_{i=1}^n \lambda_i \sin \theta_i}{\sum_{i=1}^n \lambda_i \cos \theta_i}.
\end{equation}
\end{subequations}

In our setting we have $\{\lambda_i\} = \{\lambda_\mf\}$, $\theta_0 = 0$, and $\theta_{\pm1} = \pm \theta/2$, such that after algebraic manipulations we obtain Eq.~\eqref{eq:A_all} of the main text and find for the offset phase
\begin{equation}
    \tan \theta_\offset = \frac{\lambda_0 + \Delta \lambda \sin(\theta/2)}{\lambda_0 + (\lambda_{-1} + \lambda_{+1}) \cos(\theta/2)}.
\end{equation}

A pointwise phase reconstruction is possible if all $\lambda_\mf$ are known and can be obtained from
\begin{align}
\begin{split}
\label{eq:theta_all}
    \cos(\theta/2) =& \frac{\lambda_0 (\Lambda - \lambda_0)}{\Delta \lambda^2 - (\Lambda - \lambda_0)^2} \\%
    &\pm \sqrt{\frac{\Delta \lambda^2 \bigl[ \Delta \lambda^2 + 2 \Lambda \lambda_0 - \Lambda^2 \bigr]}{\bigl[ \Delta \lambda^2 - (\Lambda - \lambda_0)^2 \bigr]^2} %
    - \frac{(\amp_\all / \amp_0)^2}{\Delta \lambda^2 - (\Lambda - \lambda_0)^2} },
\end{split}
\end{align}
where we introduced $\Lambda = \sum \displaylimits_\mf \lambda_\mf$.

\subsection{Histogram Fitting Routine}
We generate all histograms by binning the $300$ measurements of each signal into $\sqrt{300} \cong 18$ bins.
The width of the bins for each $T$ setting is adapted so that all $18$ bins cover the whole range of acquired data and therefore depends on the signal's maximum value.
This change of bin size can be seen in Figs.~\ref{fig:3}(\textbf{B}) and \ref{fig:4}.
The square root rule is an established rule for creating histograms with an adequate resolution in our setting, although we note that this is not a trivial matter and leaves plenty of room for refinement \cite{shimazaki:2007:bins, lohaka:2007:bins}.

\subsubsection{Probability Density Function}
We start from the signal form of the main text, \ie{} $S =\base + \amp \cos (\phi_0 + \theta_\offset)$, where $\amp$ is a constant amplitude, $\phi_0$ is a (potentially fluctuating) phase that is used to scan a fringe, $\theta_\offset$ is a fixed phase, and $\base$ is a normally distributed random variable describing baseline fluctuations.

At the same time, fluctuations of $\phi_0$ translate into a PDF for the observed amplitude~\cite{chattamvelli:2021:arcsine}
\begin{equation}
    f_\amp(a) =     \begin{cases}
                        1/\Bigl(\pi\sqrt{\amp^2 - a^2}\Bigr) \!\!\!\!\!\! & , \, -\amp < a < \amp\\
                        0 & , \, \text{else}
                    \end{cases},
\end{equation}
if one assumes a uniform phase distribution with support $[0, 2\pi)$.
By linearly scanning $\phi_0$ in our experiment, we ensure that this uniform phase distribution is a good approximation, even in the absence of phase fluctuations, \ie{} for short $T$.
However, any underlying PDF of $\phi_0$ can be used to adapt $f_\amp$.

Moreover, we assume that baseline fluctuations are independent and identically distributed, such that their PDF is given by a normal distribution
\begin{equation}
    f_\base(b) = \frac{1}{\sigma \sqrt{2 \pi}} \exp \bigg[ - \frac{(b - \mu)^2}{2 \sigma^2} \bigg],
\end{equation}
with mean $\mu$ and variance $\sigma^2$, according to the central limit theorem.
The PDF of a sum of two random variables is the convolution $\ast$ of their respective PDFs, \ie{} the PDF of the signal $S$ is
\begin{equation}
\label{eq:hist_pdf}
   f_S(\mathscr{s}) = \bigl(f_\amp \ast f_\base\bigr)(\mathscr{s})= \frac{1}{\sigma \sqrt{2 \pi^3}} \int \displaylimits_{-\amp}^\amp \! \dd a \, \frac{\exp \bigg[ - \frac{(\mathscr{s} - a - \mu)^2}{2 \sigma^2} \bigg]}{\sqrt{\amp^2 - a^2}} .
\end{equation}
For $\amp \ge 2\sigma$, this PDF shows a symmetric double-peak distribution.
These maxima are roughly separated by $2 \amp$, their width is determined by $\sigma$, and the PDF rapidly decays for $|\mathscr{s}|\rightarrow \infty$.
Contrary, for $\amp < 2\sigma$ the two maxima start to overlap, causing the double-peak structure to become washed out.

\subsubsection{Initial Guesses}
\label{subsubsec:init_guess}
For fitting the histogram's PDF, \cf{} Eq.~\eqref{eq:hist_pdf}, we use \verb|scipy.optimize.curve_fit| from the \verb|scipy| package in \verb|python|~\cite{virtanen:2020:scipy}, returning $\amp_\text{fit}$, $\sigma_\text{fit}$, and $\mu_\text{fit}$.
To guarantee reliable convergence of the fit, accurate initial parameter guesses are essential.
\textit{Nota bene}: The \verb|scipy.optimize.curve_fit| method is based on a gradient approach which is computationally efficient, however, the PDF $f_S$ is not differentiable with respect to $\amp$ even in a distributional sense.
We noticed issues related to this non-differentiability when using poor initial guesses.

The initial guess of the histogram's mean $\mu_\text{guess}$ is given by explicitly calculating the signal's mean.
A reliable initial guess for the standard deviation of the baseline fluctuations $\sigma_\text{guess}$ can be extracted from the rapid decay $|\mathscr{s}|\rightarrow \infty$ of the PDF:
First, we determine one maximum of the histogram at $\mathscr{s}_\text{max}$.
Starting from this maximum, we scan outward across the histogram to locate the position $\mathscr{s}_k = k \mathscr{s}_\text{max}$ with $0 < k < 1$, where the maximum has dropped to the fraction $k$.
We then roughly approximate the rapid outward drop of the histogram by an exponential decay, resulting in
\begin{equation}
    \sigma_\text{guess} = \lvert \mathscr{s}_\text{max} - \mathscr{s}_k \rvert \sqrt{-1/(2 \ln k)},
\end{equation}
which is well-defined since $\ln k < 0$.
The choice $k=1/4$ gave robust and reliable initial guesses.

The initial guess of the amplitude $\amp_\text{guess}$ is obtained from the above guesses via
\begin{equation}
    \amp_\text{guess} = \lvert \mathscr{s}_\text{max} - \mu_\text{guess} \rvert + \sigma_\text{guess}.
\end{equation}
We numerically confirmed that $\mathscr{s}_\text{max} \cong \mu \pm (\amp - \sigma)$, up to numerical factors.
An analytical form is only possible for $\sigma=0$, where we find $\mathscr{s}_\text{max} = \mu \pm \amp$ as expected.

\subsubsection{Fitting Procedure}
When fitting a single histogram, we make use of the initial guesses according to Sect.~\ref{subsubsec:init_guess}.
In the state-selective case, we have access to $S_{\pm1}$, $S_\summ$, and $S_\diff$, such that we can perform four histogram fits.
We start by fitting  the histogram's PDF to $S_{\pm1}$ based on the initial guesses from the previous section, returning $\amp_{\pm1, \text{fit}}$, $\sigma_{\pm1, \text{fit}}$, and $\mu_{\pm1, \text{fit}}$.
We set $\amp_{0}$ to be the mean $(\amp_{+1, \text{fit}}+\amp_{-1, \text{fit}})/2$.
Next, we use the mean $(\sigma_{+1, \text{fit}}+\sigma_{-1, \text{fit}})/2$ as the initial guess for $\sigma_{\summ, \text{fit}}$ and $\sigma_{\diff, \text{fit}}$, unless it is smaller than any bin width of $S_{\pm1}$.
In the latter case, we use the initial guess routine of Sect.~\ref{subsubsec:init_guess} for $\sigma_{\summ, \text{fit}}$ and $\sigma_{\diff, \text{fit}}$.
Both $S_\summ$ and $S_\diff$ fluctuate with the same standard deviation $\sigma$, identical to that of $S_{\pm1}$, because for $\base_i \sim \mathcal{N}(\mu_i, \sigma_i^2)$ the random variable $Z = \lambda_1 \base_1 + \lambda_2 \base_2$ is also normally distributed, \ie{}  $Z \sim \mathcal{N}(\lambda_1 \mu_1 + \lambda_2 \mu_2, \lambda_1^2 \sigma_1^2 + \lambda_2^2 \sigma_2^2)$.
Since we have $\sigma_{1,2} = \sigma$ and $\lambda_{\pm1} = 1/\sqrt{2}$, we find $\lambda_1^2 \sigma_1^2 +\lambda_2^2\sigma_2^2 = \sigma^2$.

Last, we restrict the parameter $\amp_{\summ, \text{fit}}$ to $[0, \sqrt{2} \amp_{0, \text{fit}}]$, but otherwise estimate the amplitudes as in Sect.~\ref{subsubsec:init_guess}.
These limits ensure the correct connection between $S_{\pm1}$ and $S_\summ$, allowing for a phase reconstruction via Eq.~\eqref{eq:A_sum}.

We apply the same procedure to the difference signal, but use $S_\diff= \sqrt{2}\amp_0|\sin(\theta/2)|$ for phase reconstruction.
Unlike the amplitude and standard deviation of the baseline fluctuations, the mean is not an issue.
Therefore, we use the mean of the respective signal as the initial guess.

\subsection{Ellipse Fitting Routine}
As a comparison to PEAC, we deploy an established approach to evaluate correlated signals based on fitting an ellipse to a bivariate parametric plot of two signals.
We assume that the signals of the $\mf=\pm1$ substates have the parametric form $S_{\pm1}(\phi_0) = \base_{\pm1}+\amp_{\pm1} \cos(\phi_0 \pm \theta/2) $ with amplitudes $\amp_j$, baselines $\base_j$, common phase $\phi_0$, and differential phase $\theta$.
The bivariate data span ellipses in the $(S_{+1},S_{-1})$ plane, where the eccentricity of the ellipse encodes the differential phase.
We translate the parametric form into the algebraic description $0 = c_{+1}^2 S_{+1}^2 + c_{-1}^2 S_{-1}^2 + c_{0}^{\vphantom{2}} S_{+1}^{\vphantom{2}} S_{-1}^{\vphantom{2}} + d_{+1,-1}^{\vphantom{2}} S_{+1}^{\vphantom{2}} + d_{-1,+1}^{\vphantom{2}} S_{-1}^{\vphantom{2}} + d_0^{\vphantom{2}}$ of conic sections with the coefficients
\begin{align}
\begin{split}
\label{eq:ell_coeff}
        c_j  & = (\amp_j \sin  \theta)^{-1} \quad \text{and} \quad
        c_{0}  = -2 c_{+1} c_{-1} \cos \theta \\
        d_{i,j}  &= -2 \base_i^{\vphantom{2}} c_i^2 - \base_j^{\vphantom{2}} c_{0}^{\vphantom{2}} \quad \text{and} \\
        d_0 & = \base_{+1}^2 c_{+1}^2+\base_{-1}^2 c_{-1}^2 + \base_{+1}^{\vphantom{2}} \base_{-1}^{\vphantom{2}} c_{0}^{\vphantom{2}} -1    
\end{split}
\end{align}
which for ellipses have to satisfy $c_0^2 - 4 c_{+1}^2 c_{-1}^2 = - 4 c_{+1}^2 c_{-1}^2 \sin^2\theta < 0$ for $\theta \in (0, \pi) + \pi \mathds{Z}$.
For $\theta=\pi \mathds{Z}$ the ellipse degenerates onto straight lines in the case of perfect correlation or anti-correlation.
Direct evaluation of Eq.~\eqref{eq:ell_coeff} gives access to the differential phase
\begin{equation}
\label{eq:theta_ell}
    \theta = \pm \arccos \Big( -c_0/\sqrt{4c_{+1}^2 c_{-1}^2 } \Big) + \pi \mathds{Z}.
\end{equation}

As mentioned in the main text, diagonalising the quadratic form $c_{+1}^2 S_{+1}^2 + c_0^{\vphantom{2}} S_{+1}^{\vphantom{2}} S_{-1}^{\vphantom{2}} + c_{-1}^2 S_{-1}^2$ and assuming equal amplitudes $A_{+1}=A_{-1}$ implying $c_{+1}=c_{-1}$ leads to the two principal axes of the ellipse along the $\bigl(1, \sgn(c_0)\bigr)^\text{T}/\sqrt{2}$ and $\bigl(1, -\sgn(c_0)\bigr)^\text{T}/\sqrt{2}$ direction of the bivariate coordinate space, \ie{} the diagonal and anti-diagonal depending on the sign of $c_0$.

Expressed in the bivariate plane, the signal for $\amp_{\pm1}= A_0$ is
\begin{equation}
\label{eq:signals_bivariate}
    \begin{pmatrix}
        S_{-1} \\ S_{+1}
    \end{pmatrix}
    =
    \begin{pmatrix}
        B_{-1} \\ B_{+1}
    \end{pmatrix}
    +  A_0 \cos \frac{\theta}{2}
    \begin{pmatrix}
        1 \\ 1
    \end{pmatrix}
    \cos \phi_0
    +  A_0 \sin\frac{\theta}{2}
    \begin{pmatrix}
        1 \\ -1
    \end{pmatrix}
    \sin \phi_0.
\end{equation}
Clearly, this equation describes an ellipse with semi axes along the diagonal/anti-diagonal of the bivariate plane.
Hence, we perform a geometric fit to general form
\begin{equation}
    \begin{pmatrix}
        S_{-1} \\ S_{+1}
    \end{pmatrix}
    =
    \begin{pmatrix}
        B_{-1} \\ B_{+1}
    \end{pmatrix}
    + a
    \begin{pmatrix}
        \cos \xi \\ \sin \xi
    \end{pmatrix}
    \cos \phi_0
    + b
    \begin{pmatrix}
        - \sin \xi \\ \cos \xi
    \end{pmatrix}
    \sin \phi_0,
\end{equation}
where $a,b$ are the moduli of the semi axes and $\xi$ is the ellipse's rotation angle in the bivariate plane.
Using Eq.~\eqref{eq:sum_of_cosines_all} to cast this into the form of Eq.~\eqref{eq:signals_bivariate} gives
\begin{equation}
    \theta = \arctan \biggl( \frac{b \sin \xi}{a \cos \xi} \biggr) - \arctan \biggl( \frac{-b \cos \xi}{a \sin \xi} \biggr).
\end{equation}
This form of the differential phase $\theta$ allows using the $2$-argument arctangent of many programming languages.
Minimising the sum of the squared Euclidean distance gives rise to the so-called geometric ellipse fits~\cite{gander:1994:ellipses, ridley:2024:investigation, corgier:2025:optimized}.
We use \verb|scipy.optimize.least_squares| and \verb|scipy.optimize.minimize_scalar| for this minimisation process with the physical bounds $B_{\pm 1} \in [-1 ,1]$, $a, b \in [0 , \sqrt{2}]$, and $\xi \in [-\pi, \pi]$ as we work with normalised signals.
Due to its computational low cost, we use the Hal\'{\i}\v{r}--Flusser algorithm~\cite{halir:1998:numerically, halir_flusser:2025:ellipse} to fit the algebraic ellipse equation~\eqref{eq:ell_coeff} to our data as an initial guess for the computationally more demanding geometric ellipse fits.

\subsection{Bootstrapping}
We assume that the experimental uncertainties of the homoscedastic interferometric signals $S$ stem from fluctuations of the signal's baseline $\base$, parametrised by the standard deviation $\sigma$ of a normal distribution in our model, while the amplitude of the signal is assumed to be constant.
Since for any interferometer time $T$ we fit the PDF to all recorded data points, no uncertainty of this procedure can be extracted in a direct way.
To estimate the uncertainties of measured quantities, we therefore implement a bootstrapping approach~\cite{efron:1986:bootstrap,efron:1994:introduction,efron:2000:bootstrap}, assuming our data provide a representative sample of the underlying probability distribution.
For each interferometer time $T$, we use all $300$ available data points and randomly draw a set of $300$ data points with replacement.
In addition to the original data, we generate $999$ random sets, giving us access to $1000$ independent datasets per $T$, each comprising $300$ points.
These datasets are then used to obtain the quantities of interest such as the amplitude $\amp$ or the differential phase $\theta$ via least-square fitting methods.
Based on the extracted values, we compute mean and standard deviations for each quantity, which are reported in the main body of this article.

To validate the reliability of the bootstrapping procedure, we compare it against randomly generated numerical datasets.
We find no substantial difference in the standard deviations between random and bootstrapped datasets.

\subsection{Numerical Replication}
To investigate properties such as the bias $\theta_\bias$ or the uncertainty $\Delta\theta$ with finer resolution than experimentally being accessible with reasonable effort, we resort to a numerical replication of our experiment.
For the numerically generated data, we use the amplitude $A_{0,\text{exp}} = 0.824$ and baseline fluctuations $\sigma_\text{exp} = 0.063$ extracted from the experimental datasets, setting the mean baseline offset $\mu=0$, since it corresponds to a translation that can be neglected.
Similar to bootstrapping the experimental data, we produce $1000$ numerical datasets containing $300$ randomly generated values $S_{\pm 1}$ for each phase $\theta$ of interest.
Here, we markedly increase the sampling of $\theta$ compared to the experiment to give access to finer details.
Each random value of $S_{\pm 1}$ is computed by sampling a random phase $\phi_0\in[0,2\pi)$ and a baseline value $\base_{\pm 1}$ drawn from a normal distribution with mean $\mu=0$ and variance $\sigma_\text{exp}^2$.
We implement the random number generation using the \verb|numpy.random.default_rng.uniform| and \verb|numpy.random.default_rng.normal| functions of the \verb|numpy| package in \verb|python| \cite{harris:2020:array,python:2025:python3}.
By producing $1000$ individual datasets, rather than bootstrapping a single set, as it was done with the experimental data, we can check whether a systematic effect is apparent in the bootstrapping procedure itself, which assumes that the experimental sample size is sufficient to describe the underlying PDF.
In our case, we find no systematic difference between the bootstrapped experimental data and numerically generated random data, hence we observe no systematic effects.

\subsection{Origin of a Bias in PEAC}
We identified three main challenges that cause a bias in PEAC:
(i) the signal's amplitude $\amp$ becomes comparable to the standard deviation of the baseline fluctuations $\sigma$;
(ii) the true amplitude is close to the fit boundaries;  and
(iii) resolution limits imposed by bin size.
In contrast to ellipse fits at the points of degeneracy, where the algebraic structure reduces to a line and fitting data to an ellipse is ill-posed, the underlying PDF used in PEAC is correct for all differential phases.
Thus, the observed bias is not intrinsic to PEAC, but reflects limitations of any fit procedure in regimes of ambiguous parameters, parameters close to boundaries, or a finite resolution.

Addressing (i):
Due to the amplitude modulation, the PDF of the histogram transitions from a double-peak to a single-peak distribution, as the two maxima begin to overlap until only one maximum remains, depending on the differential phase.
Although no analytical form exists for this transition, we numerically found a threshold at $\amp / \sigma \cong 1.7777$.
Below this limit, two equally plausible fits are possible because of the ambiguity in the combination of $\amp$ and $\sigma$.
This ambiguity leads to a wider spread in the amplitude values $\amp$, resulting in a wider spread of $\theta_\recons$ and by that increasing its uncertainty $\Delta \theta$.
Consequently, the true amplitude value becomes less likely, resulting in a distorted distribution overestimating the mean.
A maximum likelihood estimation~\cite{pezze:2025:joint} instead or a Bayesian inference strategy~\cite{stockton:2007:bayesian, pelluet:2025:atom, barret:2015:correlative} could in principle be used to mitigate this issue.

Addressing (ii):
To ensure consistency between $S_{\pm1}$ and $S_\summ, S_\diff$, we impose the fit limits $\amp_\summ, \amp_\diff \in [0, \sqrt{2} \amp_0]$.
For example, the upper limit guarantees that a phase reconstruction via Eq.~\eqref{eq:A_sum} is possible, as higher amplitudes imply values outside the domain of the $\operatorname{arccos}$ function.
However, this constrain leads to a truncated distribution of fit parameters, whose mean underestimates the true amplitude value.
Over- and underestimating is a general limitation of any fit routine with physical or logical bounds of the fit parameters.

Addressing (iii):
We use a fixed number of $18$ bins, but the range of signal values $S_\summ$ and $ S_\diff$ changes with the differential phase $\theta$ because of the modulation of $\amp_\summ$ and $\amp_\diff$.
A wider range of values implies bigger bin sizes reducing the fit resolution.
In combination with (ii), this effect is most likely the reason for the bias in the difference signal (along the major axis) being bigger than the bias in the sum signal (along the minor axis) at $\theta = \pi$.

\vspace{-3mm}
\begin{acknowledgments}
We thank S. Abend, F. Di Pumpo, A. Friedrich, S. Kanthak, M. Meister, and D. Reinhard for stimulating discussions and suggestions.
We acknowledge contributions from the Terrestrial Very-Long-Baseline Atom Interferometry (TVLBAI) proto-collaboration.

\noindent\textbf{Funding:}\\
The QUANTUS-VI project is supported by the German Space Agency at the German Aerospace Center (Deutsche Raumfahrtagentur im Deutschen Zentrum für Luft- und Raumfahrt, DLR) with funds provided by the Federal Ministry for Economic Affairs and Climate Action (Bundesministerium für Wirtschaft und Klimaschutz, BMWK) due to an enactment of the German Bundestag under grant no. 50WM2450E (QUANTUS-VI).
The grant recipient is EG, and this funding supports the position of DD.
\end{acknowledgments}

\noindent\textbf{Author Contributions:}\\
DD and DP contributed equally to this work.\\
Conceptualisation: EG, GB, DD (supporting), DP (supporting)\\
Data curation: DD, DP, LL (supporting)\\
Formal analysis: DD, DP, LL (supporting)\\
Funding acquisition: EG, GB\\
Investigation: DD, DP, LL (supporting)\\
Methodology: DD, DP, EG (supporting), GB (supporting)\\
Software: DD, DP, LL (supporting)\\
Supervision: EG, GB\\
Validation: DD, DP, LL, EG, GB\\
Visualisation: DD, DP, EG (supporting), GB (supporting)\\
Writing -- original draft: DD, DP, EG (supporting)\\
Writing -- review \& editing: DD, DP, EG, GB, LL (supporting)

\noindent\textbf{Competing Interests:}\\
All authors declare that they have no competing interests.

\noindent\textbf{Data, Code, and Materials Availability:}\\
All data and code needed to evaluate and reproduce the results in the paper are present in the paper, the Supplementary Materials and a publicly accessible online repository at \url{https://www.doi.org/10.5281/zenodo.21340342} based on \url{https://github.com/TU-Darmstadt-APQ/PEAC.git}.
This study did not generate new materials.
\bibliography{references}

\clearpage

\onecolumngrid

\section*{Supplementary Materials}
\setcounter{section}{0}
\setcounter{subsection}{0}
\setcounter{equation}{0}
\setcounter{figure}{0}
\setcounter{table}{0}
\renewcommand{\thesection}{S\arabic{section}}
\renewcommand{\theequation}{S\arabic{equation}}
\renewcommand{\thefigure}{S\arabic{figure}}
\renewcommand{\thetable}{S\arabic{table}}
\section{Geometric vs. Algebraic Ellipse Fits}
\begin{figure}[!h]
    \centering
    \includegraphics[scale=.9]{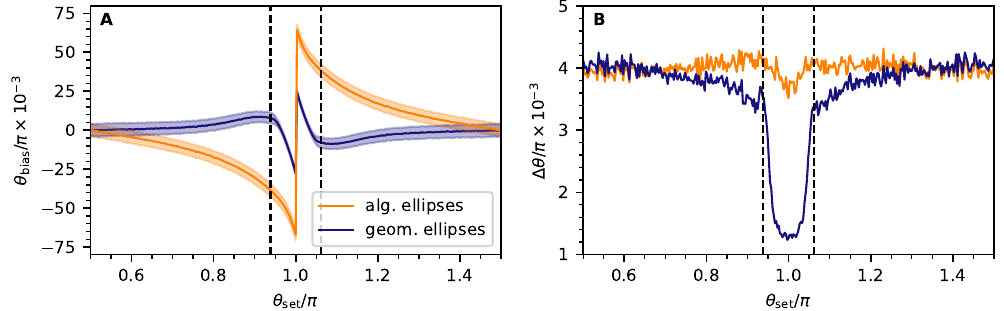}
    \caption{\textbf{Comparison of geometric and algebraic ellipse fits.} (\textbf{A}) Fitting ellipses using the geometric description (dark blue) reduces the bias $\theta_\text{bias}$ compared to the algebraic fit (orange) based on the Hal\'{\i}\v{r}--Flusser algorithm~\cite{halir:1998:numerically,halir_flusser:2025:ellipse}. 
    Nevertheless, independent of the fitting routine, ellipses become bias free only for $\theta=\pi/2$ and show an increased bias at the degeneracy point $\theta=\pi$.
    (\textbf{B}) In contrast to the bias, the uncertainty $\Delta\theta$ remains similar for both fitting routines outside the interval marked by dashed lines.
    The apparent reduction of the bias inside this region stems from the fit converging to the boundary of the major semi axis at $\sqrt{2}$, which effectively reduces the uncertainty. 
        }
    \label{fig:S1}
\end{figure}

In the main body of the article, we compared PEAC to geometric ellipse fits, realised by a least-squares implementation, for more details see Materials and Methods. 
Since both signals $S_{+1}$ and $S_{-1}$ are normalised, the semi axes of the corresponding ellipses have to be confined to the interval $[0,\sqrt{2}]$, a physical constraint that we implemented in our fitting routine.
To optimise the computational cost, we used the standard approach of fitting the algebraic form of the ellipse via the algebraic Hal\'{\i}\v{r}--Flusser algorithm~\cite{halir:1998:numerically,halir_flusser:2025:ellipse} and convert its output to the geometric form as an initial guess for geometric ellipse fitting.
Naturally, it can be expected that geometric ellipse fits return a lower bias and uncertainty when compared to the algebraic form~\cite{gander:1994:ellipses, ridley:2024:investigation}. 
To verify this established result, we compare both approaches with special emphasis on then degeneracy point $\theta=\pi$.
Figure~\ref{fig:S1} shows bias (\textbf{A}) and uncertainty (\textbf{B}) for both algebraic (orange) and geometric (dark blue) ellipse fits in the interval $\pi/2\leq\theta\leq3\pi/2$. 
As expected, the geometric ellipse fits generate a lower bias over the whole interval with comparable uncertainty, and reduce it by a factor of approximately two at the degeneracy point $\theta=\pi$.
Nevertheless, both fitting techniques are only bias free when the ellipse turns into a circle at $\theta=\pi/2$. 
Inside the region marked by dashed lines, where PEAC has an increased bias and uncertainty due to the ambiguity of fit parameters, geometric ellipse fits show a reduced uncertainty, as in this region the fitting routine trends towards the boundary condition $\sqrt{2}$ for the major half axes of the ellipse, effectively fixing one parameter at this value. 
Hence, the reduction in uncertainty is deceptive and does not provide an increase in accuracy, especially given the substantial bias in this region. 
In summary, we confirm the expected performance gain of geometric ellipse fits compared to the algebraic approach at the cost of increased computation time.

\clearpage

\section{Impact of Baseline Fluctuations}
To generalise the results of the main body of the article beyond situations tailored to our specific experimental parameters, we examine the impact of various baseline fluctuations.
In analogy to the numerical replication of the experiment, we generate a set of random bivariate data scanning the differential phase $\theta_\sett \in [\pi/2, \pi]$ for different baseline fluctuations $\sigma$, and the intrinsic amplitude $A_0=0.824$.

\begin{figure}[!h]
    \centering
    \includegraphics[scale=.9]{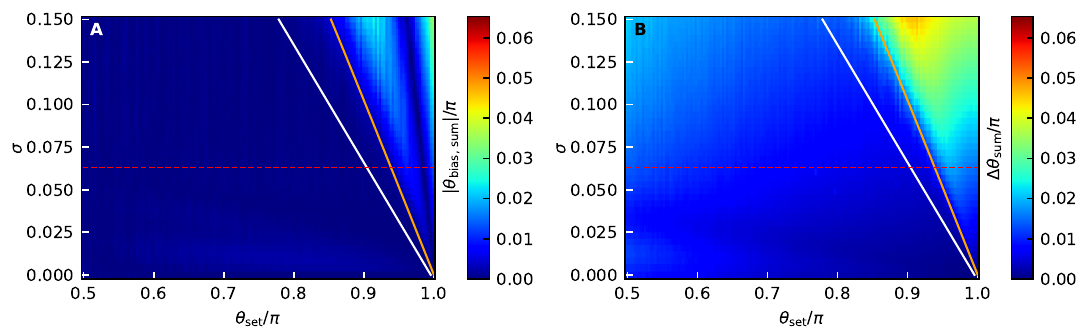}
    \caption{\textbf{Baseline-noise scan for bias and uncertainty of PEAC.}
        The orange line indicates the criterion $A \cong 1.78 \sigma$, while the white line highlights the working point of PEAC.
        The horizontal red dashed line at $\sigma=0.063$ marks our experimental noise value.
        (\textbf{A})~Absolute value of the bias for PEAC along $S_\summ$.
        (\textbf{B})~Phase uncertainty for PEAC along $S_\summ$.
    }
    \label{fig:S2}
\end{figure}

\begin{figure}[!h]
    \centering
    \includegraphics[scale=.9]{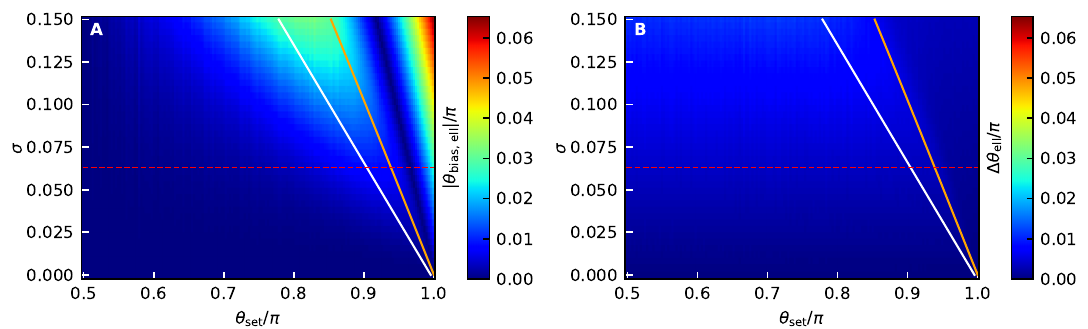}
    \caption{\textbf{Baseline-noise scan for bias and uncertainty of ellipse fitting.}
        The orange line indicates the criterion $A \cong 1.78 \sigma$, while the white line highlights the working point of PEAC.
        The horizontal red dashed line at $\sigma=0.063$ marks our experimental noise value.
        (\textbf{A})~Absolute value of the bias for ellipse fitting.
        (\textbf{B})~Phase uncertainty for ellipse fitting.
    }
    \label{fig:S3}
\end{figure}

We employ PEAC along $S_\summ$ and, for comparison, geometric ellipse fits to these numerical datasets.
The results of these parameter scans are shown as density plots in Figs.~\ref{fig:S2} for PEAC and \ref{fig:S3} for geometric ellipse fits, respectively.
In both figures, panel~(\textbf{A}) shows the modulus of the bias and panel~(\textbf{B}) the corresponding phase uncertainty.
For reference to the experimental parameters used in the main body of the article, we include a horizontal red dashed line indicating the experimental baseline fluctuations of $\sigma = 0.063$.
As discussed in the article, the transition from a two-peak to a one-peak PDF occurs at $A \cong 1.78 \sigma$.
This threshold is indicated in Figs.~\ref{fig:S2} and \ref{fig:S3} by the orange line and supports the claim that the ambiguity of the fit parameters $A$ and $\sigma$ leads to a bias and increased uncertainty in PEAC.

In the given parameter region, we can linearise the $A \cong 1.78 \sigma$ criterion, establishing a linear connection between the value of $\theta_\sett$ at which the transition occurs and $\sigma$, as confirmed by the orange line.
The orange line essentially marks the region of substantial bias and uncertainty increase for PEAC along $S_\summ$.
In contrast, ellipse fits show a substantially larger bias over a broad range of $\theta_\sett$, extending beyond the region bounded by the orange line.
Nevertheless, the phase uncertainty of PEAC is larger than that of ellipse fitting, especially in the region around degeneracy where ellipses underestimate the systematic error induced by bias.

Additionally, we indicate with the white line the working point $\theta_\text{wp}$, \ie{} the phase of minimal phase uncertainty in PEAC.
This working point always falls within the region of negligible bias, providing an uncertainty comparable to ellipse fits while maintaining reduced bias.
By fitting a line to the minimum, we find that this behaviour can be approximated by $\theta_\text{wp}/\pi \cong 1 - 1.45 \sigma$ within our parameter regime.
This trend is in agreement with the expectation that $\theta_\text{wp} \rightarrow \pi$ for $\sigma \rightarrow 0$.

Moreover, we observe an advantage of PEAC at higher noise levels, in particular for our experimental value, while ellipses perform the best in low-noise environments.
Thus, we used our numerical model to explore PEAC from high- to low-noise environments, highlighting its complementarity to ellipse fitting.
These insights are supported by the criterion $A \cong 1.78 \sigma$, which allows the identification of the effective bias-free region and the working point of PEAC.

\section{Complementarity of PEAC and Ellipse Fitting}
\label{sec:complementarity}
\begin{figure}[t]
    \centering
    \includegraphics[scale=.9]{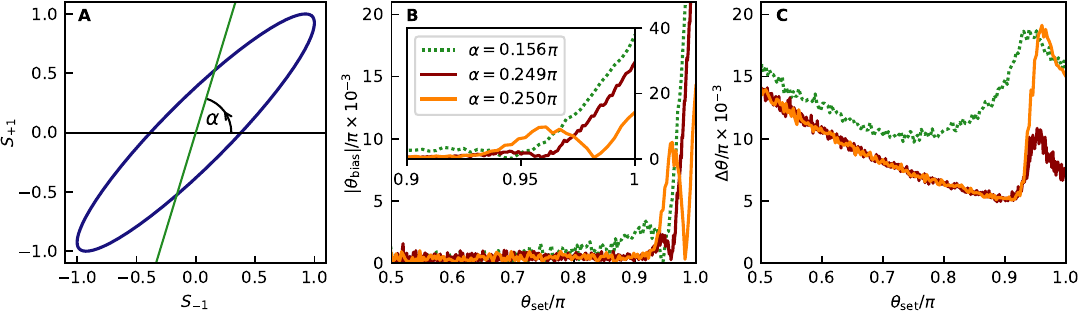}
    \caption{
    \textbf{Rotated Projection Along the Ellipse. }%
    (\textbf{A}) Schematic depiction of the projection through the data parametrised by the rotation angle $\alpha$ in the bivariate plane.
    Along this angle, the data is projected to obtain a histogram and to perform  an analysis based on PEAC.
    Choosing $\alpha=\pi/4$ corresponds to the analysis performed with $S_\summ$.
    (\textbf{B}) Resulting magnitude of the bias $\theta_\bias$ for different choices of $\alpha$.
    Increasing deviation from $\alpha=\pi/4$ leads to larger bias, whereas values in the vicinity of $\alpha=\pi/4$ show an extended bias-free region.
    Nevertheless, near degeneracy, $\alpha=\pi/4$ remains the favourable operating direction.
    (\textbf{C}) Uncertainty $\Delta \theta$ of reconstructed phase for different $\alpha$.
    Increasing deviation from $\alpha=\pi/4$ induces a substantial increase of the phase uncertainty.
    In contrast, values close to $\alpha=\pi/4$ maintain the same minimal uncertainty at the working point, but exhibit reduced uncertainty closer to degeneracy compared to $\alpha=\pi/4$.
    Nevertheless, for $\alpha=0.249 \pi$, the bias exceeds the uncertainty, leaving $\alpha = \pi/4$ as the preferable choice near degeneracy.
    }
    \label{fig:S4}
\end{figure}

If correlated data in the bivariate plane have been recorded, ellipse fits can be performed as described in the main body of the article.
However, such data also allows PEAC to be applied in addition to, or in combination with, this analysis.
PEAC is flexible in that, beyond the semi-axes along the diagonal and anti-diagonal, any projection angle $\alpha$ can be used for parameter extraction, as sketched in Fig.~\ref{fig:S4}(\textbf{A}).

In fact, considering the general rotated signal $S(\alpha)$ with amplitude $A(\alpha) = A_0 \sqrt{1 + \sin(2\alpha) \cos \theta}$ (\cf{} Eq.~(3) of the main text for $\lambda_0=0, \lambda_{-1}=\cos \alpha, \lambda_{+1}=\sin \alpha$), we can calculate its phase uncertainty for $\alpha \neq \pi/2 \ \mathds{Z}$
\begin{equation}
\label{eq:theta_unc_wrt_alpha}
    \Delta \theta(\alpha) = \frac{2 \Delta A(\alpha)}{A_0} \frac{\sqrt{1+\sin(2\alpha) \cos \theta}}{|\sin(2\alpha) \sin\theta|},
\end{equation}
where $\Delta A(\alpha)$ is the uncertainty of the rotated oscillation amplitude.
Assuming that the variance $\Delta A$ is in\-de\-pendent of $\alpha$, the phase uncertainty for $\theta \in (\pi/2, 3\pi/2)$ becomes minimal for $\alpha = \pi/4$ implying $\lambda _{\pm1}=1/\sqrt{2}$, \ie{} an equal superposition of the signals.
In this sense $S_\summ$ is optimal since $\Delta \theta(\alpha) \ge \Delta \theta(\alpha=\pi/4)$.
Under this assumption, the ellipse's principal axes $S_\summ$ and $S_\diff$ are the directions one should choose when employing PEAC for $\theta \in (\pi/2, 3\pi/2)$ and $\theta \in (3\pi/2, \pi)$, respectively.

Nevertheless, this analysis neglects the treatment of $\Delta A(\alpha)$, which in practice depends on the employed fitting routine as discussed in Sect.~2.3 of the main text.
For this reason, we performed a numerical analysis by scanning the angle $\alpha$ as shown in Fig.~\ref{fig:S4}(\textbf{A}), along which we generate histograms and perform PEAC.
%
Here, we show in detail specific values of $\alpha$ allowing to assess PEAC's performance close to $\alpha=\pi/4$ and far away.
In Fig.~\ref{fig:S4}(\textbf{B}) and (\textbf{C}) we present bias and uncertainty, respectively. 
As expected from Eq.~\eqref{eq:theta_unc_wrt_alpha}, increasing deviations from $\alpha=\pi/4$ lead to larger bias and uncertainty.
For values in the vicinity of $\pi/4$ we observe a behaviour similar to that along the semi axes.
However, compared to $\alpha = \pi/4$, the onset of the bias occurs at smaller deviations from the degeneracy point and the uncertainty is even reduced in the vicinity of degeneracy.
This effect can be understood as follows:
By rotating around $\alpha$, the two peaks of the PDF are projected on an axis where their separation is increased, such that the transition from a two-peaked to a single-peaked PDF occurs closer to degeneracy.
Consequently, the fit routine can profit from a reduced ambiguity in $A$ and $\sigma$ when fitting the histogram's PDF.
Exactly this benefit can be observed for $\alpha$ in vicinity, of but not at, $\pi/4$, as illustrated by the choice $\alpha = 0.249\pi$.

While other work on quantum clocks claims to gain orders of magnitude better uncertainty ~\cite{zhou:2025:geometric} when working close to, but not at $\alpha=\pi/4$, in our case, we do not observe a comparable improvement with PEAC.
In addition, although the onset of the bias occurs closer to degeneracy, the magnitude of its variation is larger, as shown by the inset of Fig.~\ref{fig:S4}(\textbf{B}).
Crucially, $\theta_\bias$ exceeds the statistical uncertainty $\Delta\theta$.
As a consequence, $\theta_\recons \pm \Delta \theta$ does not include the true value and phase estimates cannot be trusted through the entire $\theta$ range.
In contrast, an equal superposition with $\alpha=\pi/4$, \ie{} $S_\summ$, has a smaller bias close to degeneracy but an increased uncertainty so that the true value is always included within the uncertainty band of the estimated phase.
The criterion $A< 1.78 \sigma$, together with the freedom to choose $\alpha$ arbitrarily, allows optimising the analysis for a given experimental implementation.
Thus, whenever correlated bivariate data are recorded, PEAC allows choosing $\alpha$ for an improved analysis tailored to the specific needs of the specific estimation task.
These findings highlight the complementarity of PEAC to pure ellipse-fitting approaches.

\FloatBarrier

\subsection{Full Data Scan and Normalised Bias}
Here, we present the complete data demonstrating PEAC's flexibility in choosing the rotation angle $\alpha$.
This freedom allows us to investigate the effect of population imbalances.
We generate numerical datasets in which the parameters of the PDF match the experimental conditions $A_0 = 0.824$ and $\sigma = 0.063$.
In contrast to the main article, we rely on histograms obtained by adding weighted signals $S(\alpha) = S_{-1}\cos\alpha +  S_{+1}\sin\alpha$, which corresponds to histograms projected along the angle $\alpha$.
For each angle $\alpha$, we perform a differential phase scan $\theta_\sett \in [\pi/2, \pi]$ and use PEAC to extract the bias $\theta_\bias$ and the phase uncertainty of the reconstructed phase $\Delta \theta$.

\begin{figure}[htb]
    \centering
    \includegraphics[scale=.9]{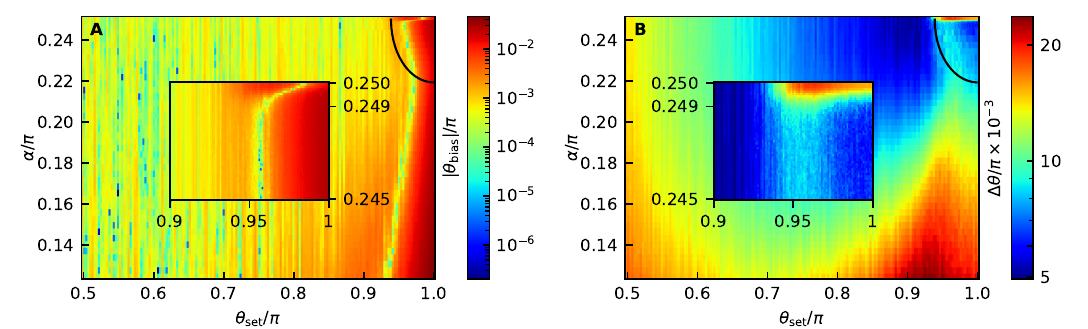}
    \caption{\textbf{PEAC in the bivariate plane. }%
        (\textbf{A}) Absolute value of the bias for PEAC in the $\alpha$ direction.
        (\textbf{B}) Phase uncertainty for PEAC in the $\alpha$ direction.
        In both panels, the black line marks the transition from a two-peak to a single-peak PDF.
        To the left of this line, the PDF exhibits two peaks, while to the right it shows a single peak.
    }
    \label{fig:S5}
\end{figure}

In the previous section, we have shown that $\alpha=\pi/4$ ($\alpha=3\pi/4$) is optimal assuming that the uncertainty $\Delta A$ is independent of $\alpha$, \ie{} $\Delta \theta(\alpha) \ge \Delta \theta(\alpha=\pi/4)$.
However, since $\Delta A(\alpha)$ may depend on the fit routine and the direction of projection, we explicitly investigate $\Delta \theta(\alpha)$ within our numerical model.
We focus on the intervals $\theta_\sett \in [\pi/2, \pi]$ and $\alpha \in [\pi/8, \pi/4]$ due to symmetries of $S_\summ$ and $S_\diff$ as well as the symmetry of exchanging $S_{-1}$ and $S_{+1}$, respectively.
The parameter scans are shown as density plots in Fig.~\ref{fig:S5}, where panel~(\textbf{A}) presents the magnitude of the bias and panel~(\textbf{B}) the phase uncertainty.
The black line marks the transition between a two-peak (left) and a single-peak (right) PDF; in the absence of such a line, the PDF is two-peaked throughout.
Where applicable, the black line is determined by $A(\alpha, \theta) \cong 1.78 \sigma$, corresponding to the generalised, $\alpha$-dependent form of the criterion $A < 1.78 \sigma$, with $A(\alpha, \theta) = A_0 \sqrt{1 + \sin(2\alpha) \cos \theta}$.
Thus, by tuning $\alpha$, one can ensure a two-peaked PDF for any differential phase $\theta$.

We observe an increase in both the bias and the uncertainty for values of $\alpha$ far from $\pi/4$.
An increase of the uncertainty for deviations from $\pi/4$ is expected from Gaussian uncertainty propagation, as detailed above, which is supported by the trend shown in panel (\textbf{B}) from top to bottom.
The uncertainty of $\Delta A(\alpha)$ becomes relevant in the vicinity of degeneracy.
Near degeneracy, the bias shows large variations and can even change sign, producing a pronounced minimum at a narrow point, while the uncertainty remains substantial.
Instead of focusing on this region, we identify the parameter regime of small bias.
For $\alpha \rightarrow \pi/4$, the inset of panel~(\textbf{A}) shows that the onset of the bias variation is closer to degeneracy for $\alpha \lesssim 0.249 \pi$.
Nevertheless, the bias at degeneracy is smaller for an equal superposition, \ie{} $\alpha = \pi/4$.

Considering the uncertainty in panel~(\textbf{B}), we find that its absolute minimum occurs at $\alpha=\pi/4$ shortly before $A(\alpha, \theta) = 1.78 \sigma$, supporting the above claim.
However, the inset shows that just before an equal superposition, the uncertainty is substantially reduced, so that already slight deviations from $\alpha=\pi/4$ lead to a reduced phase uncertainty.
In connection to the section on the impact of baseline fluctuations, where we observed that the increase in uncertainty for $\alpha=\pi/4$ depends on $\sigma$, we expect this feature in turn also to depend on the magnitude of baseline fluctuations.
In addition, for small population imbalances, the onset of bias variations shifts closer to degeneracy, accompanied by a reduced uncertainty.
This finding suggests that operating with slight population imbalances, \eg{} $\alpha=0.249\pi$, may be advantageous.
However, near degeneracy the bias variations also increase, which can introduce practical issues.
To quantifies these effects, we introduce the normalised bias $| \theta_\bias | / \Delta \theta$.

\FloatBarrier

\begin{figure}[htb]
    \centering
    \includegraphics[scale=.9]{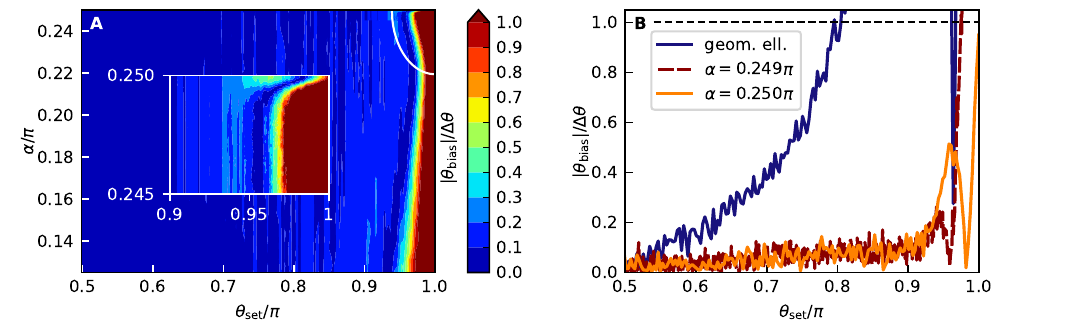}
    \caption{\textbf{Normalised bias in the bivariate plane. }%
        (\textbf{A}) Normalised bias for PEAC in the $\alpha$ direction.
        The inset shows a detailed plot of the upper right section where $0.245 \leq \alpha \leq 0.250$ and $9\pi/10 \leq \theta_\sett \leq \pi$.
        (\textbf{B}) Cuts through the density plot of the normalised bias for $\alpha=0.249 \pi$ (dark red, dashed), $\alpha=0.250 \pi$ (orange, solid), compared to the normalised bias for geometric ellipse fits (dark blue).
        In panel (\textbf{A}), the white line marks the transition from a two-peak to a single-peak PDF.
        To the left of the line, the PDF exhibits two peaks, while to the right it shows a single peak.
        In panel (\textbf{B}), the dashed black line at the top marks the threshold where $\theta_\bias/\Delta\theta = 1$. 
    }
    \label{fig:S6}
\end{figure}

The normalised bias is a measure for the quality of an estimate:
If it exceeds unity, the statistical uncertainty underestimates the systematic uncertainty, and the one-standard-deviation interval $\theta_\recons \pm \Delta \theta$ does not include the true value $\theta_\sett$.
Consequently, one ideally aims to operate in a regime where the normalised bias is less than or equal to unity.
Figure~\ref{fig:S6}(\textbf{A}) shows the normalised bias as a density plot for various angles $\alpha$ with respect to different $\theta_\sett$.
It becomes apparent that only for $\alpha = \pi/4$ the normalised bias remains less than unity (black line) over the whole range for $\theta_\sett$; otherwise, it always exceeds unity when approaching degeneracy.
In the inset of Fig.~\ref{fig:S6}(\textbf{A}) we see the behaviour of the normalised bias in detail near degeneracy and equal superpositions.

For a detailed comparison, in Fig.~\ref{fig:S6}(\textbf{B}) we also show cuts through the density plot at $\alpha=0.250\pi$ and $\alpha=0.249 \pi$, together with the normalised bias obtained from geometric ellipse fitting.
We find that the normalised bias for the geometric ellipse fit remains below unity up to $0.8\pi$ for our experimental parameters.
In contrast, PEAC performs more favourably, both due to a reduced bias and an increased phase uncertainty.
The normalised bias of PEAC for $\alpha = \pi/4$ is always smaller or close to unity throughout, such that an equal superposition does not underestimate the systematic error.
Based on absolute values of bias and uncertainty, $\alpha = 0.249 \pi$ might appear more beneficial, since the onset of bias variations shifts closer to degeneracy with a reduced uncertainty.
However, at $\alpha = 0.249 \pi$ the bias exceeds the uncertainty at degeneracy, so that the normalised bias surpasses unity.
Thus, the optimal choice of $\alpha$ depends on the specific settings and fitting routine.

\clearpage

\section{Maximum Likelihood Estimation within PEAC}
Different estimation techniques extracting the measurand from a given PDF or histogram are associated with different biases.
For example, performing a least square (LSQ) fit of the PDF to a histogram, as in the main body of the article, is sensitive to the binning of the data, which effectively reduces the resolution and introduces an additional bias.
Furthermore, the ambiguity of the fit parameters $\amp$ and $\sigma$, especially for small values of $\amp$, \eg{} close to the degeneracy point $\theta=\pi/2$, leads to an increase of bias and uncertainty.

Methods like maximum likelihood estimation (MLE) circumvent the binning of the data by directly evaluating the PDF at the measured data points through a specific likelihood function $\mathcal{L}(\hat{p}|\mathscr{s})$. 
In this \supp, we implement an MLE approach using the negative log-likelihood function 
\begin{equation}
\label{eq:nll}
    \mathcal{L}(\hat{p}|\mathscr{s})=-\sum_i \log \bigl( f_S(\hat{p}|\mathscr{s_i}) \bigr)
\end{equation}
and numerically minimise its value with respect to the parameter vector $\hat{p}$ using \verb|scipy.optimize.minimize| and \verb|scipy| package in \verb|python| \cite{virtanen:2020:scipy, python:2025:python3}.

\begin{figure}[!h]
    \centering
    \includegraphics[scale=.9]{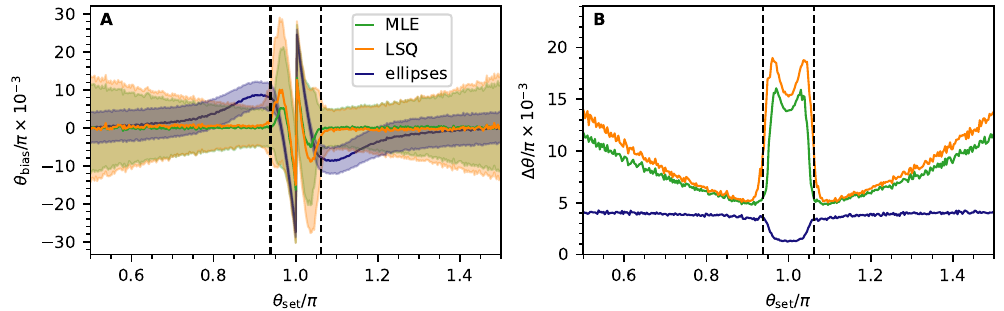}
    \caption{\textbf{Comparison of least squares (LSQ) (orange), maximum likelihood (MLE) (green), and geometric ellipse fits (dark blue).} (\textbf{A})~Bias $\theta_\text{bias}$ for differential phases $\theta_\text{set}\in\left[\pi/2,3\pi/2\right]$ and three evaluation methods. As discussed in the main article, the LSQ approach of PEAC already substantially reduces bias compared to geometric ellipse fits at the cost of an increased uncertainty. Combining PEAC with the MLE approach bypasses the need for binning the data, \ie{} calculating a histogram, and leads to a further reduction of the bias, while the qualitative trend is similar to the LSQ realisation. Inside the region marked by dashed lines, where $A < 1.78 \sigma$, all three evaluation methods show an increase in bias, zero crossings, and apparent discontinuities.
    In the case of PEAC, such effects can be partially explained by the ambiguity of the fitting parameters $A$ and $\sigma$, see the discussion in the main article.
    In contrast, the geometric ellipse fits converge at these points to the maximum value of the semi axes, \ie{} $\sqrt{2}$.
    (\textbf{B})~Uncertainty $\Delta\theta$ for the three evaluation methods.
    Similar to the bias, we observe a slight reduction of the uncertainty for the MLE compared to the LSQ approach of PEAC. As expected, the geometric ellipse fits show a smaller uncertainty, being the more precise evaluation method albeit a larger bias and thus a reduced accuracy. Inside the dashed lines, $\Delta\theta$ is even reduced for the ellipse fits, as the fit parameters converge to the upper constraint. 
        }
    \label{fig:S7}
\end{figure}

This implementation has shown a substantial increase of computational cost with a runtime of $\qty{150}{h}$ compared to the LSQ fit approach $(\qty{8}{h})$ to fit our numerical data.
In Fig.~\ref{fig:S7} we compare MLE and LSQ implementations of PEAC in bias (\textbf{A}) and uncertainty (\textbf{B}) for the summed signal $S_\summ$ of the numerical replication of the experiment. 
Additionally, we include bias and uncertainty for the geometric ellipse fits as a benchmark of standard evaluation techniques. 
Generally, we find that MLE indeed improves the performance of PEAC with a reduction both in bias and uncertainty, especially close to the degeneracy point $\theta=\pi$. 
The bias is reduced over the complete interval of $\theta$, but the gain is small compared to the substantially increased computational cost. 
Around the degeneracy point, in-between the dashed lines, the bias is reduced by a factor of two, but exactly at degeneracy, the discontinuity is of the same order of magnitude as for the LSQ approach. 
Similar to the bias, the uncertainty is slightly reduced over the complete interval with the largest improvement close to the degeneracy. 
Outside the shown interval, the uncertainty is reduced by a factor of two at even multiples of $\pi$, where one would resort to using $S_\diff$ instead. 
In conclusion, we observe an improvement of the PEAC performance when using MLE compared to LSQ at the cost of substantially increased computational time in our implementation. 

\clearpage

\section{Accuracy Assessment in Bias-Uncertainty Space}
In Fig.~\ref{fig:S7}, we present both bias and uncertainty as function of the set phase $\theta_\sett$, which can be identified as systematic and statistical errors.
In experimental data analysis, the objective is generally to provide the smallest overall error budget by balancing both systematic and statistical contributions.
However, there is no unique metric for combining these two quantities into a single performance measure:
For example, equally weighing both contributions leads to the mean squared error $\theta_\bias^2 + \Delta \theta^2$, but this metric does not distinguish between situations in which the bias exceeds the statistical uncertainty, so that the true value lies outside the error margin and the uncertainty band shown in Fig.~\ref{fig:S7}(A).

\begin{figure}[!h]
    \centering
    \includegraphics[scale=.9]{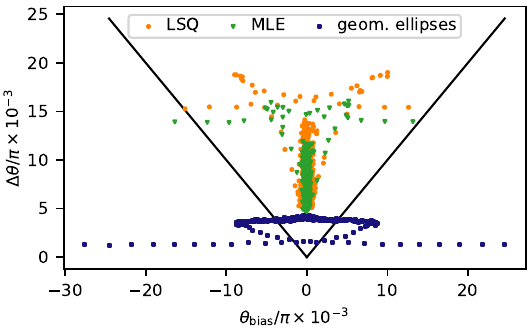}
    \caption{\textbf{Comparison of differential phase estimators in the $\theta_\bias$ and $\Delta\theta$ plane.}
    Scatter plot of bias versus uncertainty for each set phase $\theta_\mathrm{set}\in[\pi/2,3\pi/2]$ obtained from ellipse fitting (blue squares), PEAC-LSQ (yellow circles), and PEAC-MLE (green triangles).
    The solid lines denote the boundary $|\theta_\mathrm{bias}|=\Delta\theta$.
    Points inside the cone correspond to estimates whose reported uncertainty interval contains the set value and can be regarded as true estimates.
    While 100\,\% of the PEAC-LSQ results (and 99.7\,\% of the PEAC-MLE results) lie within this cone, only 62\,\% of the ellipse-fit results satisfy this criterion.
    }
    \label{fig:S8}
\end{figure}

To enable a fair comparison of PEAC-MLE, PEAC-LSQ, and ellipse fitting, we avoid combining trueness, quantified by the bias, and precision, quantified by the standard deviation, into a single performance metric.
Instead, Fig.~\ref{fig:S8} presents the overall error budget by plotting the uncertainty of each data point as a function of its bias for $\theta_\mathrm{set}\in[\pi/2,3\pi/2]$.
The cone enclosed by the solid lines contains all measurements for which the set value lies within the reported $1\sigma$ uncertainty interval of the estimated mean.
Estimates inside this cone therefore satisfy one criterion for true parameter estimation.
While 100\,\% of the PEAC-LSQ results (and 99.7\,\% of the PEAC-MLE results) lie within this cone, only 62\,\% of the ellipse-fit results satisfy this criterion.
The ellipse-fit points inside the cone correspond primarily to set phases located around the circular configuration, whereas the lower branch of ellipse-fit data originates from operating points close to the degeneracy.

The scatter plot also highlights the different characteristics of the estimators: 
PEAC substantially suppresses the bias and confines the results to a compact region of parameter space, whereas ellipse fits exhibit widespread bias, albeit with a generally smaller statistical uncertainty.
Finally, the uncertainties of PEAC approach the uncertainty baseline achieved by the ellipse fits.
This demonstrates that, when operating close to the working point, PEAC is competitive with ellipse fitting in all aspects, while consistently providing true estimates over all parameters and set phases of the interval.

\end{document}